\documentclass{emulateapj}
\usepackage{lscape}
\usepackage{longtable}

\begin{document}
\newcommand{\lya}{Lyman~$\alpha$}
\newcommand{\lyb}{Lyman~$\beta$}
\newcommand{\za}{$z_{\rm abs}$}
\newcommand{\ze}{$z_{\rm em}$}
\newcommand{\cmtwo}{cm$^{-2}$}
\newcommand{\nhi}{$N$(H$^0$)}
\newcommand{\degpoint}{\mbox{$^\circ\mskip-7.0mu.\,$}}
\newcommand{\kms}{\,km~s$^{-1}$}      
\newcommand{\minpoint}{\mbox{$'\mskip-4.7mu.\mskip0.8mu$}}
\newcommand{\peryr}{\mbox{$\>\rm yr^{-1}$}}
\newcommand{\secpoint}{\mbox{$''\mskip-7.6mu.\,$}}
\newcommand{\sqdeg}{\mbox{${\rm deg}^2$}}
\newcommand{\squig}{\sim\!\!}
\newcommand{\subsun}{\mbox{$_{\twelvesy\odot}$}}
\newcommand{\et}{{\rm et al.}~}
\newcommand{\msun}{\,{\rm M_\odot}}
\newcommand{\Ha}{\,{\rm H\alpha}}
\newcommand{\Hb}{\,{\rm H\beta}}

\def\ltsima{$\; \buildrel < \over \sim \;$}
\def\simlt{\lower.5ex\hbox{\ltsima}}
\def\gtsima{$\; \buildrel > \over \sim \;$}
\def\simgt{\lower.5ex\hbox{\gtsima}}
\def\arcs{$''~$}
\def\arcm{$'~$}
\def\erf{\mathop{\rm erf}}
\def\erfc{\mathop{\rm erfc}}
\title{The Direct Detection of Lyman-Continuum Emission from Star-forming
Galaxies at $z\sim 3$\altaffilmark{1}}
\author{\sc Alice E. Shapley} 
\affil{Princeton University, Peyton Hall -- Ivy Lane, Department of Astrophysical Sciences, Princeton, NJ 08544}
\author{\sc Charles C. Steidel}
\affil{California Institute of Technology, MS 105--24, Pasadena, CA 91125}
\author{\sc Max Pettini}
\affil{Institute of Astronomy, Madingley Road, Cambridge UK}
\author{\sc Kurt L. Adelberger}
\affil{McKinsey and Company, 1420 Fifth Avenue, Suite 3100, Seattle, WA 98101}
\author{\sc Dawn K. Erb}
\affil{Harvard-Smithsonian Center for Astrophysics, 60 Garden Street, Cambridge, MA 02138}

\altaffiltext{1}{Based, in part, on data obtained at the 
W.M. Keck Observatory, which 
is operated as a scientific partnership among the California Institute of Technology, the
University of California, and NASA, and was made possible by the generous financial
support of the W.M. Keck Foundation.
} 

\begin{abstract}
We present the results of rest-frame ultraviolet spectroscopic
observations of a sample of 14 $z\sim 3$ star-forming galaxies in
the SSA22a field. These spectra are characterized by
unprecedented depth in the Lyman-continuum region.
For the first time, we have detected escaping ionizing 
radiation from individual galaxies at high redshift,
with two of the 14 objects showing significant emission
below the Lyman limit. We also measured the ratio
of emergent flux density at 1500~\AA\ to that in
the Lyman-continuum region, for the individual
detections (C49 and D3) and the sample average. If a correction for
the average IGM opacity is applied to the spectra
of the objects C49 and D3, we find $f_{1500}/f_{900}$$,_{corr,C49}=4.5$ 
and $f_{1500}/f_{900}$$,_{corr,D3}=2.9$.
These numbers imply Lyman-continuum escape fractions at least
as large as that presented in \citet{steidel2001},
measured from a composite spectrum of 29 Lyman Break Galaxies (LBGs).
The average emergent flux-density ratio in our sample is
$\langle f_{1500}/f_{900},_{corr} \rangle $$=22$,
implying an escape fraction $\sim 4.5$ times lower 
than inferred from the \citeauthor{steidel2001} composite
spectrum. If this new estimate is representative
of LBGs, their contribution to the metagalactic
ionizing radiation field is $J_{\nu}(900) \sim  2.6 \times  10^{-22} 
\mbox{ erg s}^{-1}\mbox{cm}^{-2}\mbox{Hz}^{-1}\mbox{sr}^{-1}$,
comparable to the contribution of optically-selected
quasars at the same redshift. The sum of the contributions from galaxies
and quasars is consistent with recent estimates of
the level of the ionizing background at $z\sim 3$, inferred
from the H~I Ly$\alpha$ forest optical depth.
There is significant variance
among the emergent far-UV spectra in our sample, yet
the factors controlling the detection or 
non-detection of Lyman-continuum emission from galaxies
are not well-determined. Specifically, the two objects with detections
are not differentiated in a consistent manner
from the remainder of the sample in terms of their spectroscopic
or photometric properties at wavelengths longer than the Lyman limit.  
There are also differences in the average
emergent far-UV spectra of current and previous samples
used to estimate the escape fraction from star-forming
galaxies. Because we do not yet understand the source
of this variance, significantly larger samples
will be required to obtain robust constraints on the galaxy
contribution to the ionizing background at $z\sim 3$
and beyond.

\end{abstract}

\section{Introduction}
\label{sec:intro}
The metagalactic ionizing background determines
the evolution of the physical state of the intergalactic medium
(IGM). The ionization rates of H~I and He~II are 
controlled by both the amplitude of the radiation field
and its spectral shape between 1 and 4 Rydbergs. 
These properties of the ionizing background reflect the total and
relative contributions of quasars and star-forming galaxies.
It is possible to estimate the nature of the background
from the quasar proximity effect \citep{bajtlik1988,scott2000},
and also from the evolution of optical depths of H~I and 
He~II in the Ly$\alpha$ forests of both those species, which are probed by
the spectra of bright quasars \citep{shull2004,bolton2005,bolton2006,
mcdonald2001,fan2005}. Such studies indicate that, while the
comoving ionizing emissivity decreases between $z=3$ and
$z=5$, the decrease in the comoving space density of quasars
is much steeper, and that, by $z\sim 5$, quasars
cannot dominate the ionizing background \citep{fan2001,fan2005,mcdonald2001}.
Even at $z\sim 3$, the mean H~I optical
depth, and the mean and variation in the ratio 
of He~II and H~I column densities, $N(\mbox{He~II})/N(\mbox{H~I})$,
may indicate a substantial contribution from star-forming
galaxies \citep{bolton2005,bolton2006,shull2004}. 
Since Lyman-continuum radiation from galaxies appears to constitute
an increasingly important ingredient of the ionizing background at redshifts
beyond $z\sim 3$, and may dominate the radiation
field at the epoch of H~I reionization \citep{fan2001,spergel2003},
estimating the galaxy contribution directly represents a fundamental
goal for observational cosmology.

One inherent complexity in such an estimate stems from the unknown
value of $f_{esc}$, the escape fraction of ionizing photons from
the interstellar medium of a star-forming galaxy. Knowledge of this escape
fraction is required to convert the star-formation
rate density of a population of galaxies into a comoving ionizing
emissivity, a crucial component of models of reionization 
\citep{furlanetto2004,loeb2001}. Theoretical models of the
propagation of ionizing radiation through the ISM of disk galaxies
similar to the Milky Way predict an escape fraction of
$f_{esc} \leq 0.1$ \citep{dove1994,dove2000}, though this fraction
may increase for high-redshift galaxies with active 
star-formation rates, in which supernova
explosions increase the ISM porosity \citep{clarke2002}
and large-scale outflows may clear out chimneys through
which ionizing radiation can easily escape \citep{fujita2003}.
The model predictions for $f_{esc}$ depend on the detailed structure and
dynamics of the ISM, driven by the effects of supernova explosions
and stellar winds. Currently, we have only
fairly crude data on the properties of the ISM in
distant galaxies \citep{pettini2001,shapley2003}, with
minimal spatially-resolved information.  

Another approach is to measure directly the emergent Lyman-continuum
radiation from galaxies. {\it Hopkins Ultraviolet
Telescope}  and {\it Far-Ultraviolet Spectroscopic Explorer}
($FUSE$) observations of a small sample of nearby starburst
galaxies mostly yielded upper limits for $f_{esc}$
of less than 0.10 \citep{leitherer1995,hurwitz1997,deharveng2001}.
Only recently, \citet{bergvall2006} reported the first
{\it FUSE} detection of Lyman-continuum emission from a local dwarf
starburst galaxy, Haro~11, which implies $f_{esc} \sim 0.04-0.10$,
and will provide a useful low-redshift baseline for
comparing with the observations at high redshift. 

There has been some degree of controversy over the value 
of $f_{esc}$ in galaxies at $z\sim 3$. Escaping 
Lyman-continuum flux was apparently detected by \citet{steidel2001}
in a composite spectrum of 29 Lyman Break Galaxies (LBGs)
drawn from the high-redshift tail of the $z\sim 3$ LBG selection function.
If the ratio of Lyman-continuum to $1500$~\AA\ flux density observed in
this composite spectrum is typical for the entire
LBG sample, it implies $f_{esc}\sim 0.1$, 
after correction for the average IGM
opacity and the estimate that only $15-20$\% of 1500 \AA\ photons
escape from typical LBGs \citep{adelberger2000}. 
However, the spectrum may not be representative of typical LBGs, as it reflects
the mean emergent spectrum for LBGs with bluer than
average rest-frame UV colors and, therefore, objects 
less affected by dust extinction. Other attempts to
measure escaping ionizing radiation from LBGs
have yielded only upper limits. \citet{giallongo2002}
targeted two objects drawn from the bright end of the
LBG luminosity function, obtaining an upper
limit on the escape fraction that is 4 times lower
than the detection in \citet{steidel2001}. This limit
may not be as stringent if the two individual sightlines
pass through higher than average H~I optical
depth in the Lyman-continuum region -- i.e. application
of the average IGM correction factor for $z\sim 3$
may not necessarily be correct in individual cases.
\citet{fernandez2003} fit the broad-band spectral energy distributions 
(SEDs) of 27 galaxies at $1.9 \leq z \leq 3.5$, including
the effects of intrinsic Lyman-limit absorption, and
intervening Lyman series and Lyman limit absorption,
and found on average $f_{esc} \leq 0.04$. It should
be noted that there are  significant uncertainties
associated with this method, because of the use of
the broad {\it HST}/WFPC2 $F300W$ filter to probe the Lyman
limit at $z\sim 2-3$. For galaxies towards the low redshift end of
the sample, galaxy flux from above the Lyman limit
falls in this filter, while for galaxies towards the high
redshift end of the sample, the level of flux within
the filter will be modulated primarily by the effects of
IGM absorption, providing only crude information on 
the flux level just below the Lyman limit. Using
a narrow-band filter more finely tuned to probe
just below the Lyman limit for two galaxies
at $z\sim 3$, \citet{inoue2005} also found
only upper limits to the escape fraction, but
imaging with greater depth will be required
to test if the limits are in contradiction to the
\citet{steidel2001} detection.

Because of the increasing IGM optical depth
at higher redshift, it is not possible to
measure escaping ionizing radiation directly from
galaxies much beyond $z\sim 3$. Understanding
how $f_{esc}$ relates to other galaxy properties
that {\it can} be measured at $z\sim 6$
is required for assessing the contribution
of galaxies to the ionizing background
at the epoch of reionization. Therefore
it is crucial to resolve the controversy
over the escape fraction from galaxies at $z\sim 3$.
In an effort to address this question, we have
undertaken a program of deep spectroscopy
of a sample of 14 LBGs, drawn from the heart of the LBG
redshift selection function ($\langle z \rangle = 3.06$).
The mean and range of UV colors in this sample are 
representative of the total sample of LBGs, in contrast
to the bluer than average sample of \citet{steidel2001}.
Spectra were obtained with the
upgraded LRIS spectrograph on the Keck~I telescope,
which provides unparalleled sensitivity in the $3500-5000$~\AA\
wavelength range \citep{steidel2004}. The depth
in the Lyman-continuum region for individual spectra
in our sample is comparable to that in the
composite spectrum of \citet{steidel2001}.
In order to correlate the physical properties of
the ISM in these galaxies with the degree of Lyman-continuum escape,
we have also obtained sensitive observations redwards of Ly$\alpha$
for our sample, covering strong low- and high-ionization interstellar 
absorption features at twice the resolution of typical LBG
discovery spectra. 

For the first time, we present the detection of escaping ionizing radiation
for {\it individual} galaxies at $z\sim 3$. Considering our two
detections and 12 non-detections, we find significant
variation in the escape fraction from galaxy to galaxy,
and an average escape fraction which is significantly
lower than the one presented in \citet{steidel2001}.
Samples an order of magnitude larger are required to
understand the cause of variation among the emergent far-UV
spectra of LBGs in terms of stellar populations, morphology, orientation,  
and ISM physical conditions, and to generalize to the global
contribution of galaxies to the ionizing background.
In \S~2, we present our observations and reductions, while
\S~3 contains the main empirical results. In \S~4, we discuss
the emergent far-UV spectra of galaxies in our
sample, and conclude in \S~5 with the implications for
the contribution of galaxies to the ionizing background.
A cosmology
with $\Omega_m=0.3$, $\Omega_{\Lambda}=0.7$, and $h=0.7$ is assumed
throughout.

\section{Observations and Reductions}
\label{sec:obs}

\subsection{Deep Spectroscopic Observations}
\label{sec:obsobs}
We targeted galaxies in the SSA22a field
at RA=22:17:34.2, DEC=+00:15:01 (J2000)
for deep spectroscopic observations.
This field contains 146 photometric LBG candidates,
57 of which were spectroscopically confirmed
at the time the deep observations were planned \citep{steidel2003}.
Our primary targets for deep observations
were 14 bright galaxies with $23.0 \leq {\cal R} < 24.5$,
twelve of which had redshifts previously
determined from their original discovery spectra.
With the fairly high S/N observations described below,
we also measured redshifts for the remaining two galaxies.
The redshift distribution is characterized by an average
of $\langle z \rangle = 3.06 \pm 0.12$, and ranges from
$z=2.756$ to $z=3.292$. Nine of the galaxies in our
sample have redshifts that place them within the 
significant overdensity contained in the SSA22a
field at $\langle z \rangle = 3.09 \pm 0.03$ \citep{steidel1998}.
However, the SSA22a field was simply chosen in order to maximize the number
of bright $({\cal R} < 24.5)$ targets with known spectroscopic
redshifts that could all be placed on the same multi-object
slit mask, without regard for the presence of the redshift spike.

The data were obtained during
five separate observing runs in the
interval 2000 October - 2002 August, using the
Low Resolution Imaging Spectrometer \citep[LRIS;][]{oke1995}
on Keck I in its newly-commissioned double-beamed mode
\citep{steidel2004}. 
Conditions were photometric,
with seeing ranging from 0\secpoint8$-$0\secpoint9 for the majority
of the observations, and never worse than 1\secpoint0.
As there is not currently
an atmospheric dispersion corrector for LRIS, all but a few
of the exposures were taken at elevations within 30$^\circ$ of zenith
to minimize the effects of differential refraction in the blue. 
The total range in airmass spanned from 1.06  to 1.27.
Two different slit mask configurations were used in the
course of data collection, containing slits with 1\secpoint2 width.
The slit lengths varied from $\sim 10$" to $\sim 50$", with
a median of $\sim 20$". 
The masks had different sky position angles but significant
overlap in targeted objects, with 9 objects
in common between the setups. 
On the 2000 October observing run, the
slit mask had a sky position angle of $\theta = 172^{\circ}$.
The ``d560'' dichroic beam splitter was used to direct light
bluer than $\sim 5600$~\AA\ towards the
LRIS blue channel, while longer wavelength
light was sent to the red channel.
The blue light was dispersed by a 400 lines mm$^{-1}$
grism blazed at 3400~\AA, leading to a dispersion
of 1.70 \AA/pixel on the engineering-grade Tektronix
2K~$\times$~2K UV/AR-coated CCD, while the
red light was dispersed by a 600 lines mm$^{-1}$
grating blazed at 7500~\AA, yielding a dispersion
of 1.28 \AA/pixel on the red side Tektronix 2K~$\times$~2K CCD.
Most slitlets had complete
wavelength coverage spanning from the atmospheric
cutoff at 3200~\AA\ to redder than 7500~\AA.
On all subsequent observing runs (2001 June, 2001 July,
2002 June, and 2002 August) a slit mask with
a sky position angle of $\theta = -50^{\circ}$
was used. The ``d500'' dichroic
beam splitter was used to split the
incoming light beam at $\sim 5000$~\AA.
The blue light was dispersed by the same
400 lines mm$^{-1}$ grism blazed at 3400~\AA,
while the red light was dispersed by a
600 lines mm$^{-1}$ grating blazed at 5000~\AA.
For the final two observing runs (June 2002
and August 2002), the new science-grade Marconi
4K~$\times$~4K CCD mosaic was used for
the blue-side observations. The new blue
camera has smaller pixels, yielding
a dispersion of $\sim 1.07$~\AA/pixel
when used together with the 400-line grism.
For this second setup, most slitlets
had complete wavelength coverage
spanning from the atmospheric cutoff
to redder than 7000~\AA.

Our typical observing strategy consisted of
acquiring a series of 1800 second exposures
simultaneously on the red and blue sides.
The telescope was dithered slightly
($\sim 1$~arc second) between exposures
in order to sample different parts of the
detectors. We obtained 6 hours of
total integration time for both
red and blue side observations in
2000 October, $\sim 3$ hours in 2001 June,
$\sim 5$ hours in 2001 July, $\sim 2$ hours
in 2002 June, and $\sim 8$ hours in 2002 August.
Not all of the data acquired were
included in the final stacked spectra. 
For red side spectra, data from 2001 June, 2001 July,
2002 June, and 2002 August were included;
for blue side spectra, data from 2000 October, 2001
June, 2001 July, and 2002 August were included. 
At wavelengths below
4000~\AA, only 2002 August blue side data were included
in the final spectra, as the sensitivity of the science-grade
blue detector in this wavelength range
is a factor of $\sim 2$ better on average than that of the 
engineering-grade detector.\footnote{Red side data 
from 2000 October were not included because
of only partial overlap in wavelength coverage with subsequent observations,
and because of problems with background subtraction caused by 
internal reflections. Blue side data from 2002 June were not
included because the necessary twilight flat field exposures were
not obtained.} The resulting total integration
is  $\sim 8$ hours below 4000~\AA,  $\sim 22$ hours between
4000 and 5000~\AA, and $\sim 17$ hours above 5000~\AA.
For a minority of the sample,
only a subset of the this integration time was available.
Table~\ref{tab:obs} lists the ${\cal R}$ magnitudes,
$G-{\cal R}$  and $U_n-G$ colors, Ly$\alpha$ emission and interstellar
absorption redshifts, and total LRIS integration times
in the wavelength ranges $\lambda \leq 4000$~\AA, 
$4000 <  \lambda \leq 5000$~\AA\ (blue side), and $\lambda > 5000$~\AA\
(red side), for the 14 objects targeted with deep spectroscopy.

Spectroscopic flat fields for the red side data
were obtained at the end of an observing sequence
using an internal halogen lamp. While flat fields
were not used to correct blue side data taken with
the engineering-grade detector,
twilight flat fields were obtained during the August 2002
observing run using the science-grade blue detector.
The science-grade detector is characterized by a 
wavelength-dependent flatfield pattern  
that becomes especially pronounced below 4000~\AA.
Fluctuations resulting from 
imperfect correction of the flatfield pattern constitute
one of the largest sources of systematic uncertainty in our
measurements of flux in the Lyman-continuum region, as we
attempt to quantify in section~\ref{sec:resultsdetect}. 
An observation of internal arc lamps
(Hg, Ne, Ar, Kr, Xe for the October 2000
run, and Hg, Ne, Ar, Zn, Cd for all
subsequent runs) was also obtained
at the end of each observing sequence
for the purpose of wavelength calibration.
In order to flux-calibrate the deep mask spectra,
a spectrophotometric standard star was observed at the
end of each night through a long slit of width 1\secpoint0, and
with the same dichroic, grism, and grating combination as the deep
mask.

\subsection{Data Reduction}
\label{sec:obsredux}

The data were reduced using IRAF tasks,
with scripts designed for cutting
up the multi-object slit mask images into
individual slitlets, flatfielding, rejecting cosmic rays,
subtracting the background, averaging individual exposures into final summed
two-dimensional spectra, extracting to one dimension, wavelength and
flux-calibrating, and shifting into the vacuum frame. These procedures
are described in detail in \citet{steidel2003}. 
There was one substantive difference in the data reduction procedures
applied to the deep spectroscopic observations, relative to those
used for typical LBG discovery spectra. Rather
than basic redshift identification, the goals of the deep
observations include sensitive absolute measurements of
flux in the Lyman-continuum region and accurate
measurements of the depths of interstellar absorption lines
at longer wavelengths. To achieve these goals, we took considerable care with
the process of background subtraction, which affects
both of these types of measurements. 
Specifically, during the background-subtraction procedure,
we made sure to exclude from the
fit to the background all pixels within
$\pm 1$\secpoint0 of the object continuum location at each
dispersion point. This technique eliminated most of the oversubtraction
that is observed in LBG spectra with bright continua.
Also, while most slits were significantly longer than 15", 
we did not use more than this length to estimate the sky background level
in the Lyman-continuum region. A limit of 15" provided ample slit
real-estate for estimating the background level, while preventing
artifacts at large slit-distances from biasing the fit near
our targets.

The above procedures were applied to data from individual
observing runs. After the data from each run were fully processed
to one-dimensional, wavelength-calibrated, flux-calibrated,
vacuum-frame spectra, we combined the spectra
from different observing runs for individual objects.
For a given object, the spectra obtained on different
observing runs ranged in S/N by as much as a factor of
$\sim 3-4$, with the 2002 August spectra providing the
highest S/N. Therefore, in constructing the final summed
spectrum for each object at wavelengths longer than
4000~\AA,  we weighted the spectrum
from each observing run by the square of its S/N. As described
earlier, below 4000~\AA, only data acquired in 2002 August with
the science-grade detector were included in the final spectra.
The final S/N per resolution element in the continuum redwards of Ly$\alpha$
ranges from $\sim 5-25$ for our sample, with a median value of $\sim 12$.
With 1\secpoint2 slits, the resolution was determined by the
seeing and resulting object sizes, which we measured
along the slit spatial dimension to be smaller than
the slit-width in almost all cases. In the final combined spectra,
we estimate typical resolution elements of 
$\sim 4$~\AA\ and  $\sim 7$~\AA\ on the red and blue sides, respectively.

For 9 of the 14 galaxies, both
Ly$\alpha$ emission and average interstellar absorption
redshifts were measured, whereas only interstellar
absorption redshifts were measured for the 5
galaxies with no detectable Ly$\alpha$ emission.
Three galaxies were characterized by double-peaked
Ly$\alpha$ emission. 
The average offset between Ly$\alpha$ emission and
interstellar absorption redshifts for this sample
is consistent with what is observed in a much
larger sample of LBGs \citep{shapley2003}.
While the current sample of spectra represent a significant improvement
over typical LBG discovery spectra, in terms of
signal-to-noise, they are still not deep enough for the robust
measurement of the wavelengths of weak stellar photospheric features,
and therefore direct systemic redshifts.
Instead, galaxy systemic redshifts were estimated with the formulae
presented in \citet{adelberger2003}, 
based on stronger interstellar features,
and used to shift the spectra into the rest frame.

\section{Results}
\label{sec:results}

\subsection{Direct Detection of Lyman-Continuum Emission}
\label{sec:resultsdetect}

\begin{figure*}
\plotone{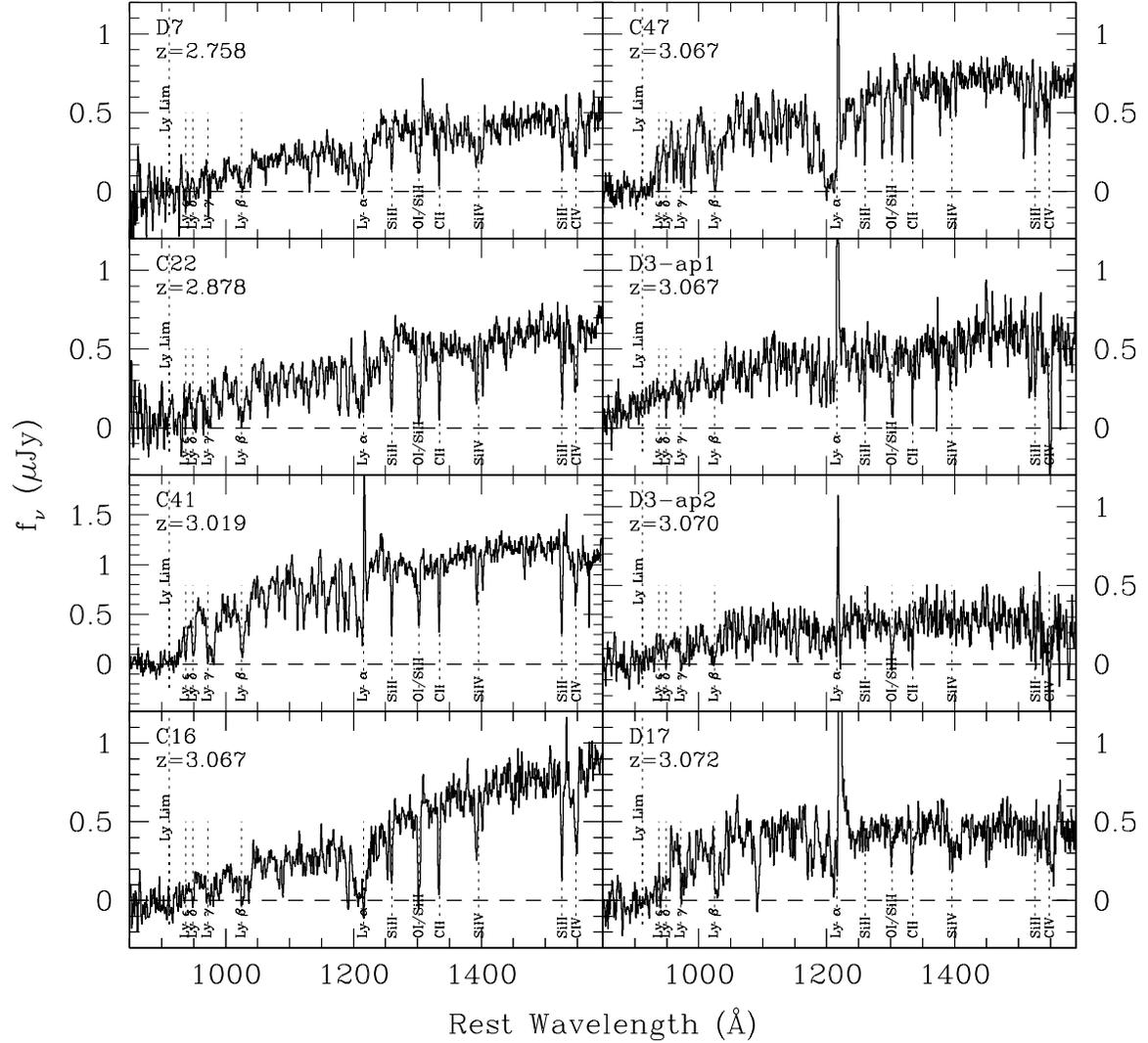}
\caption{Deep LRIS spectra of LBGs in the SSA22a field. The spectra
have been shifted into the rest frame, and the flux units
are $\mu$Jy. The strongest spectral features have been indicated
in each panel, including Ly$\alpha$ emission/absorption, absorption
from Ly$\beta$ and higher-order Lyman series, and from low- and high-ionization
interstellar lines. The Lyman limit is also marked. In this
panel, the spectrum of D3-ap1 is the only one in which significant
Lyman-continuum emission was detected. Typical exposure times in
the Lyman-continuum region are 8 hours using the
UV-optimized LRIS-B setup, and the 1-$\sigma$ uncertainty on
the average flux density in the $880-910$~\AA\ wavelength range
(due to pixel-to-pixel fluctuations) is
$\sim 1\times 10^{-31}\mbox{ erg s}^{-1}\mbox{cm}^{-2}\mbox{Hz}^{-1}$.
}
\label{fig:panel1}
\end{figure*}

\begin{figure*}
\plotone{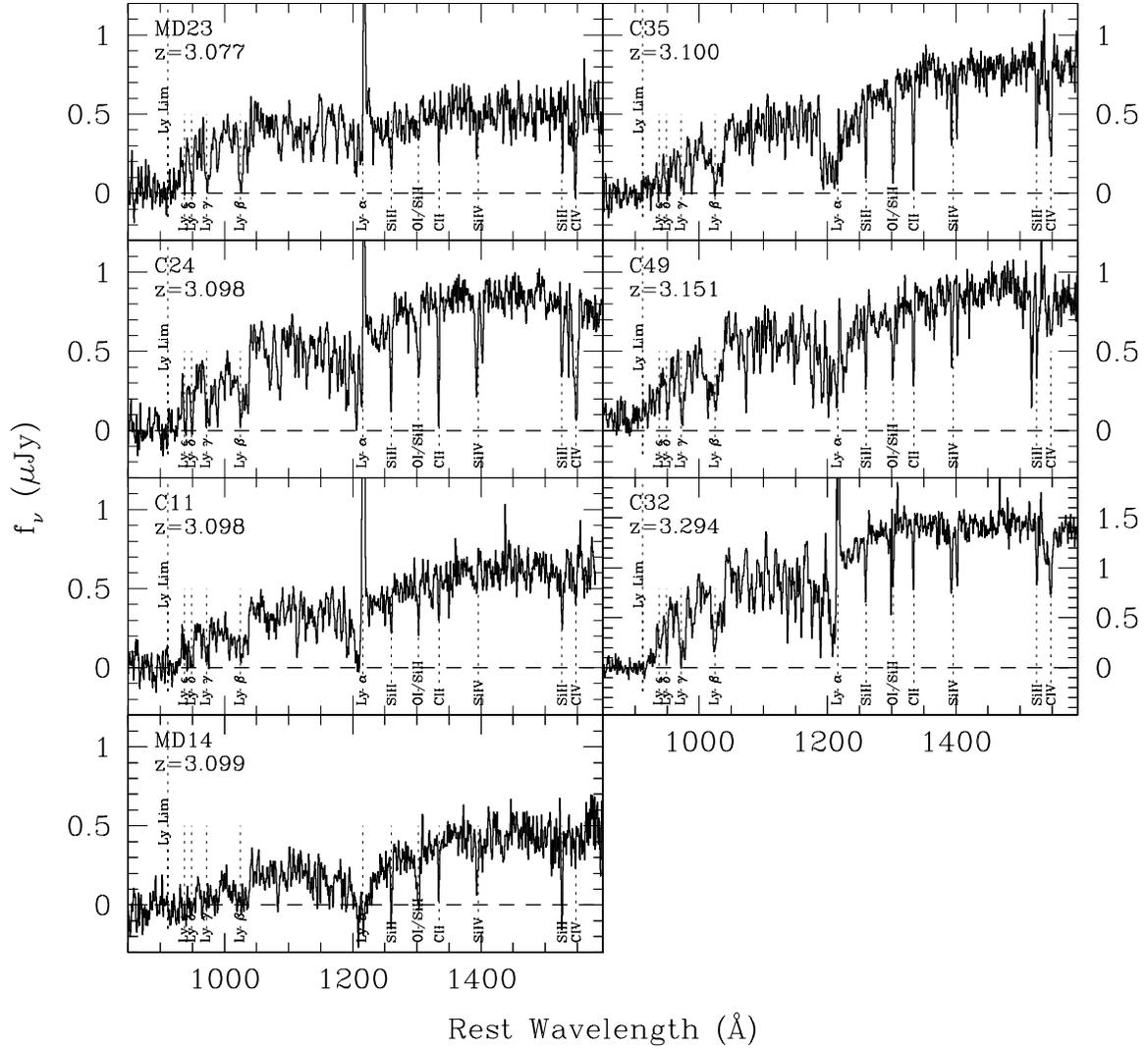}
\caption{Deep LRIS spectra of LBGs in the SSA22a field (continued).
Units, labels, and exposure times as in Figure~\ref{fig:panel1}.
In this panel, the spectrum of C49 is the only one in which significant
Lyman-continuum emission was detected.
}
\label{fig:panel2}
\end{figure*}

Figures~\ref{fig:panel1} and \ref{fig:panel2} show 
rest-frame ultraviolet spectra for the sample over 
the rest-frame wavelength range $800-1600$~\AA. 
Blue and red side spectra have
been joined at the dichroic wavelength, which
ranges in rest-frame wavelength from 1175~\AA\ for the
highest redshift object in the sample to 1345~\AA\ for
the lowest redshift object. In each spectrum,
the most prominent spectral features have been marked. The strongest
feature is H~I Ly$\alpha$, 
which appears in a variety of forms in our sample. The Ly$\alpha$
profiles range from strong, complete absorption 
($W_0(\mbox{Ly}\alpha) \sim -40$\AA) to a mixture
of emission and absorption ($W_0(\mbox{Ly}\alpha) \sim 0$\AA), 
to strong emission ($W_0(\mbox{Ly}\alpha) \sim 30$\AA). The large
range of Ly$\alpha$ profiles and equivalent widths in
$z\sim 3$ LBGs has been examined previously \citep{shapley2003}, and may
provide important clues about the properties of the 
ISM in these galaxies. In an effort to showcase 
interstellar metal absorption lines and the extremely faint
Lyman-continuum region, while placing all of the objects on similar
vertical scales, we truncated the full vertical scale of
Ly$\alpha$ for the six objects with the largest Ly$\alpha$
emission equivalent widths (D3-ap1, D17, MD23, C24, C11, C32).
In addition to Ly$\alpha$, we mark low-ionization
interstellar absorption lines from Si~II $\lambda 1260$, 
O~I $\lambda 1302$ + Si~II $\lambda 1304$,
C~II $\lambda 1334$, and Si~II $\lambda 1526$,
and high-ionization features from Si~IV $\lambda\lambda 1393,1402$
and C~IV $\lambda\lambda 1548,1550$. The Si~IV 
and C~IV features also contain contributions from stellar wind
absorption and emission. Finally, the H~I Lyman limit at
912~\AA\ is indicated.
We note that two spectra are included for the object, D3 (D3-ap1 and D3-ap2).
This object is characterized by two distinct components
separated by 200~\kms\ in velocity, and 1\secpoint9 on the sky
($\sim 15$ proper kpc at $z=3.07$). 

\begin{figure*}[t!]
\plotone{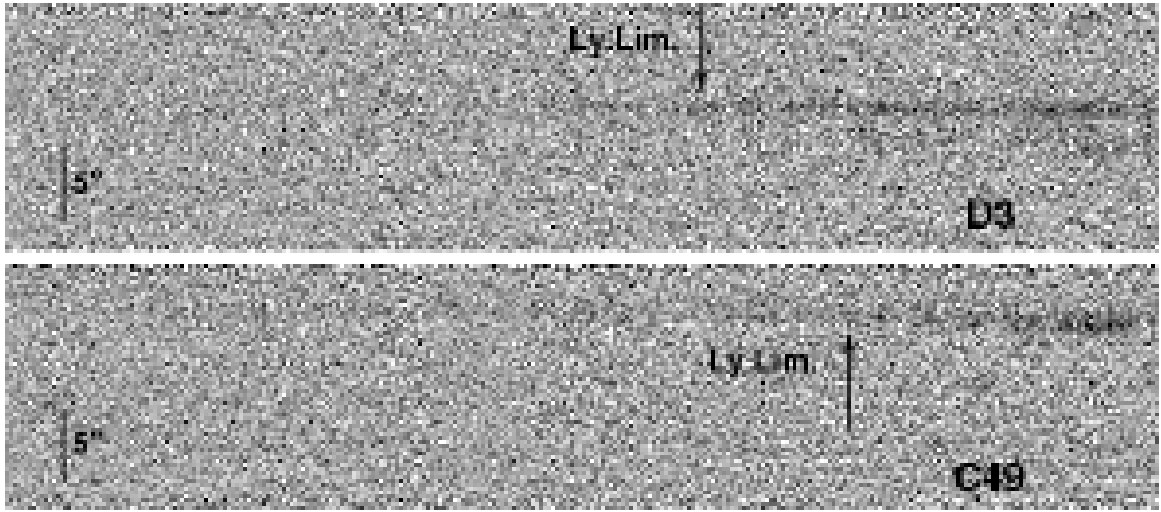}
\caption{Two-dimensional spectra of D3 and C49,
objects with Lyman-continuum detections.
Wavelength increases from left to right,
and the spatial scale along each
slit is indicated with a vertical bar of 5" in extent.
The displayed portions of the spectra
span in observed wavelength from $3350-3950$~\AA,
corresponding to $\sim 830-970$~\AA\
and $\sim 810-950$~\AA\ in the rest frames of D3 and C49,
respectively. In each two-dimensional spectrum,
the continuum clearly extends below the observed
wavelength of the Lyman limit, which is indicated with
an arrow. D3 actually consists of two components,
as is apparent at the long-wavelength (right-hand)
edge of this spectrum. However, the lower component (D3-ap2),
exhibits no significant flux in the Lyman-continuum range.
}
\label{fig:lycdet2d}
\end{figure*}

\begin{figure*}
\epsscale{0.6}
\plotone{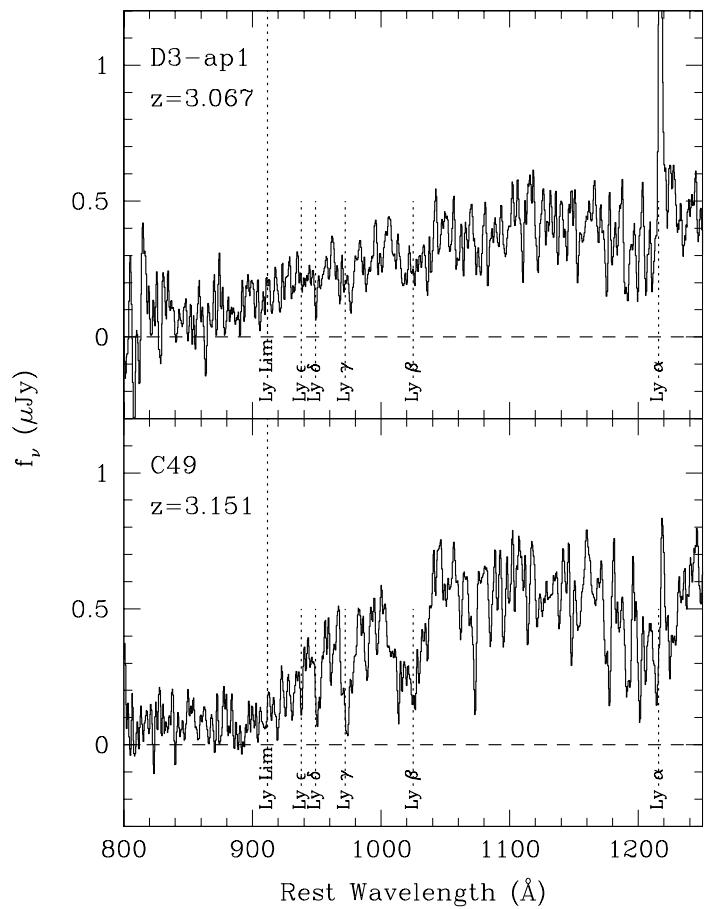}
\epsscale{1.0}
\caption{Zoomed-in one-dimensional spectra of D3 and C49,
objects with Lyman-continuum detections. The region
extending down from Ly$\alpha$ through the Lyman-continuum
region is featured.  Multiple higher-order Lyman-series are significantly
detected in C49.
The weakness of Ly$\beta$ absorption
in D3-ap1 is striking, however, as is the lack of a clear break below
the Lyman limit. We note that
Haro~11, the nearby starburst galaxy with
a recent Lyman-continuum detection, is also characterized
by the lack of a strong break at 912~\AA\ in the rest-frame
\citep{bergvall2006}.  Motivated by the $z\sim 3$ average ionizing
photon attenuation length of $\Delta z\simeq 0.18$
\citep{madau1999}, we have generally evaluated
the level of flux in the Lyman-continuum region over the
narrow range of $880-910$~\AA\ in this work. However, both
D3-ap1 and C49 exhibit significant Lyman-continuum emission
well below 880~\AA.
}
\label{fig:lycdet}
\end{figure*}

\clearpage

The measurement of escaping ionizing radiation constituted
one of the principal objectives in obtaining such deep spectra
of $z\sim 3$ LBGs. In order to measure the average flux 
density directly below the Lyman limit, $f_{900}$, we 
adopted the strategy presented in \citet{steidel2001}.
The rest-frame wavelength range used for evaluating
$f_{900}$ was $880-910$~\AA, chosen to match theoretical estimates
of the mean free path, $\Delta z \simeq 0.18$, for
Lyman-continuum photons at $z\sim 3$ 
\citep{madau1999}. In general, the evaluation of $f_{900}$ over a 
wavelength baseline significantly longer than the mean free
path of $\sim 30-40$~\AA\ in the rest frame would be more indicative of the
effects of opacity from intervening intergalactic systems than
from the intrinsic ISM of the galaxies.

Closer inspection of Figures~\ref{fig:panel1}
and \ref{fig:panel2} reveals significant Lyman-continuum flux in
two spectra, those of D3-ap1 ($z=3.067$), and C49 ($z=3.151$).
For D3-ap1, we find 
$f_{900} = 11.8 \pm 1.1 \times 10^{-31} 
\mbox{ erg s}^{-1}\mbox{cm}^{-2}\mbox{Hz}^{-1}$,
while for C49, we measure
$f_{900} = 6.9 \pm 1.0 \times 10^{-31} 
\mbox{ erg s}^{-1}\mbox{cm}^{-2}\mbox{Hz}^{-1}$.
The 1-$\sigma$ errors listed here correspond to statistical,
pixel-to-pixel, fluctuations.
In both of these objects, the detection of flux below
the Lyman limit is significant enough to
stand out in two-dimensional, unextracted spectra,
which are shown in Figure~\ref{fig:lycdet2d}. A zoomed-in
one-dimensional view of the Lyman-continuum regions is 
provided in Figure~\ref{fig:lycdet}. In both the two-dimensional
and zoomed-in one dimensional spectra, it is apparent that
leaking flux exists at a roughly constant level
well past 880~\AA, down to $\sim 810$~\AA\
in D3-ap1, and $\sim 800$~\AA\ in C49.  As mentioned above, the
object D3 consists of two distinct continuum components. 
While the component D3-ap1 exhibits significant flux at wavelengths shorter
than 912~\AA, D3-ap2 has no corresponding flux, indicating the complex
nature of this source, and the limits on the physical size of the region
where Lyman-continuum emission is escaping.
In the 13 remaining spectra (12 additional objects, plus D3-ap2),
there are no significant detections of emission below the Lyman limit.
Among individual spectra in our sample,
the average 1$\sigma$ uncertainty in $f_{900}$ is 
$\sigma_{900,stat} = 1.15  \times 10^{-31}
\mbox{ erg s}^{-1}\mbox{cm}^{-2}\mbox{Hz}^{-1}$, which
corresponds to $m_{AB}=28.75$,
and a 3$\sigma$ upper limit of $m_{AB}=27.56$. 
These numbers, again, only represent pixel-to-pixel fluctuations. 

\begin{figure}
\plotone{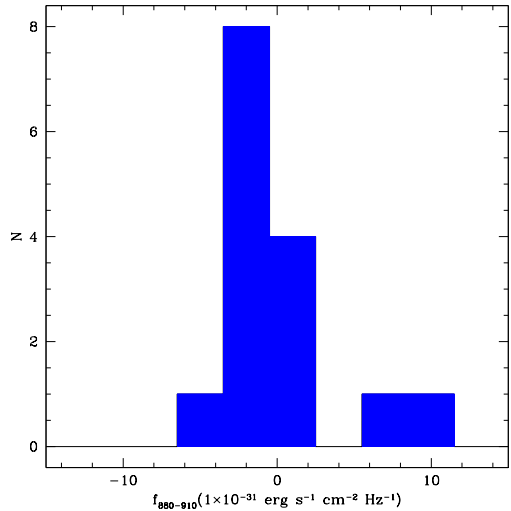}
\caption{Sample distribution of flux densities in the Lyman-continuum
wavelength range, 880-910~\AA. The two significant detections, for
C49 and D3-ap1, are clearly offset from the remainder of
the sample. Objects without significant detections have flux densities
distributed around an unphysical, negative value, which
is significantly different from zero when considered as the average
of 13 non-detections. To ensure that the objects
without significant detections are distributed around zero,
we applied a systematic additive correction to all of
the spectra, at the level of $\Delta f_{900} =   1.3  \times 10^{-31}
\mbox{ erg s}^{-1}\mbox{cm}^{-2}\mbox{Hz}^{-1}$.
As the objects without
significant detections may also be distributed around
a value that is slightly positive, the correction we applied
represents a lower limit.
The 1$\sigma$ width of the distribution for objects
without significant detections indicates
the total level of uncertainty on an individual measurement
of $f_{900}$, including contributions from both pixel to pixel and
systematic errors.
}
\label{fig:lycont_hist}
\end{figure}

As shown in Figure~\ref{fig:lycont_hist}, 
upon closer inspection we find that the average flux density 
at $880-910$~\AA\ for the objects
without significant detections is distributed around a value
that is slightly, but significantly, negative.
This artifact most likely reflects difficulty in achieving
good near-UV flatfields, free from systematic spatial fluctuations.
Such systematic undulations are visible in the two-dimensional
spectra of some of the objects without significant Lyman-continuum detections.
In order to correct for this systematic effect, we make the assumption 
that the objects without significant detections should 
have Lyman-continuum flux densities
distributed around zero, and add a constant value of 
$\Delta f_{900} =   1.3  \times 10^{-31}
\mbox{ erg s}^{-1}\mbox{cm}^{-2}\mbox{Hz}^{-1}$
to all spectra in the sample. 
Accordingly, the Lyman-continuum flux densities listed
above for C49 and D3-ap1 also include this systematic correction.
The offset applied represents a lower-limit, 
merely the amount required to make the non-detections
distributed around zero flux, rather than an unphysical,
negative value. We cannot 
rule out the possibility that the individual non-detections
are distributed around a small, slightly positive value, and therefore
require a larger correction, but for the remainder of the discussion
we will adopt the conservative lower-limit to interpret the data.
To place the described correction in context,  
we note that it represents on average
2\% of the continuum level at 1500~\AA\ for objects in our sample.

The width of the distribution of Lyman-continuum
flux densities shown in Figure~\ref{fig:lycont_hist} indicates
the total level of uncertainty in the individual $f_{900}$ values, including
both pixel-to-pixel uncertainties and systematic sources
of error arising from sky subtraction and flat-fielding.
This $1\sigma$ width is $1.9\times 10^{-31}
\mbox{ erg s}^{-1}\mbox{cm}^{-2}\mbox{Hz}^{-1}$. Given
the typical value of the $f_{900}$ pixel-to-pixel uncertainty,
$\sigma_{900,stat}$, we find that the contribution to the
error from systematic sources is 
$\sigma_{900,sys}\sim 1.5 \times 10^{-31} 
\mbox{ erg s}^{-1}\mbox{cm}^{-2}\mbox{Hz}^{-1}$.
This value pertains
to $f_{900}$ values measured from individual spectra
and must be divided by the square-root of the number
of spectra for the associated composite spectra.
Throughout the remainder of the current work, we separately
indicate {\it statistical} pixel-to-pixel, and {\it systematic} flat-fielding-
and sky-subtraction-related errors.

\begin{figure*}[t!]
\plotone{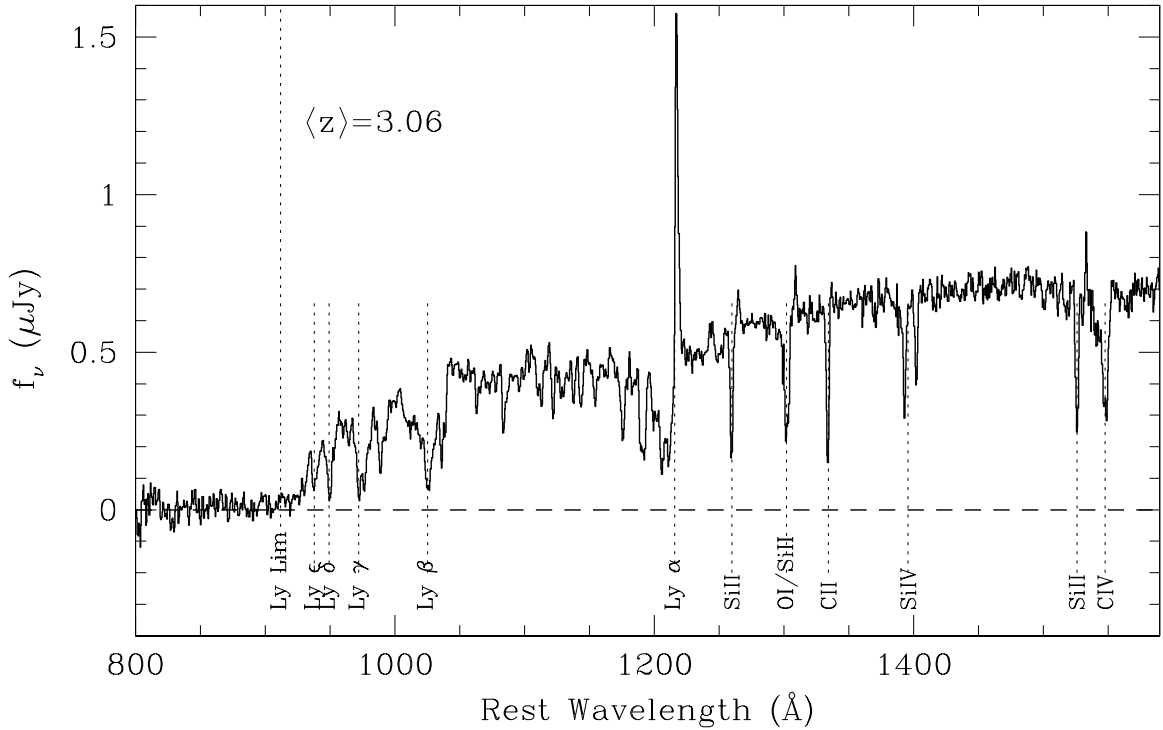}
\caption{Composite spectrum of deep LBG spectra.
This plot represents the average of the sample of
spectra displayed in Figures~\ref{fig:panel1} and \ref{fig:panel2},
and shows flux in the Lyman-continuum region that is formally
significant at greater than a 3$\sigma$ level.
The sample average ratio of $1500$~\AA\ to
Lyman-continuum flux density, uncorrected for IGM opacity,
is $\langle f_{1500}/f_{900} \rangle = 58 \pm 18_{stat} \pm 17_{sys}$.
}
\label{fig:bsysrspec}
\end{figure*}

As well as considering the Lyman-continuum flux
from individual objects in our deep sample, we also measured the properties
of a composite spectrum constructed from the whole sample,
shown in Figure~\ref{fig:bsysrspec}. This spectrum
represents a simple average of all of the individual spectra
following their systematic correction as described above.
Given the manner in which the spectra
were corrected, the mean $f_{900}$ represents
the contributions from C49 and D3-ap1 alone, averaged
over the whole sample.
In the average composite spectrum, we find 
$f_{900} = 1.08 \pm 0.28_{stat} \pm 0.37_{sys}\times 10^{-31}
\mbox{ erg s}^{-1}\mbox{cm}^{-2}\mbox{Hz}^{-1}$. 
While it does not reflect the total uncertainty in the Lyman-continuum
flux-density level, the pixel to pixel
noise represents a 1$\sigma$ depth of $m_{AB} = 30.3$,
almost a factor of 3 deeper than the corresponding
$m_{AB}=29.1$ 1$\sigma$ depth of the composite
LBG spectrum used to measure Lyman-continuum emission 
in \citet{steidel2001}. 

\subsection{Properties of Galaxies with Lyman-Continuum Detections}
\label{sec:resultD3C49}
We now consider the observed properties
of D3 and C49 in order to discern
if they stand out from objects without
significant Lyman-continuum flux detections.
There are a number of additional observations
of D3, including ground-based and {\it HST/NICMOS} 
near-IR imaging, ground-based
near-IR spectroscopy, and deep narrow-band imaging tuned
to the wavelength of Ly$\alpha$ at the redshift of the
SSA22a galaxy overdensity. There were no spectroscopic
observations of C49 prior to the deep ones presented here, and
there are accordingly fewer multi-wavelength data points for
this object. Figure~\ref{fig:specpanel} includes plots of several different
rest-frame UV properties for objects
in the deep sample, including Ly$\alpha$ equivalent width;
average low-ionization interstellar absorption equivalent width
(based on the strengths of Si~II $\lambda 1260$, 
O~I $\lambda 1302$ + Si~II $\lambda 1304$,
C~II $\lambda 1334$, and Si~II $\lambda 1526$);
$E(B-V)$; and rest-frame UV absolute magnitude, uncorrected
for dust extinction. D3 and C49 are highlighted in each
figure, as well as the average values for the three different LBG samples
listed in Table~2. 

\subsubsection{Rest-frame UV Properties}
\label{sec:resultD3C49uv}
D3 has the brightest rest-frame UV luminosity in the
current sample, with ${\cal R}_{AB}=23.37$. The flux in
the ${\cal R}$-band
represents the sum of the flux from both components of D3
(i.e. D3-ap1 and D3-ap2), which were treated as a single
object in the original photometric catalog.
As estimated from the ratio of continuum levels in
the spectra of D3-ap1 and D3-ap2, D3-ap1 is approximately
two times as bright as D3-ap2, and this ratio is roughly
constant over the wavelength range $4000-6300$~\AA,
where the flux-calibration is accurate. 
The ratio between the D3-ap1 and D3-ap2 spectra should closely
resemble the true component ratio, 
since the slit position angle of $-50^{\circ}$ east
of north is within $10^{\circ}$ of the observed angle
between the two components in the optical imaging data. 
C49 is brighter than the median of the deep sample
in rest-frame UV luminosity, while not as extreme
as D3. In a broader comparison of rest-frame UV luminosity, D3 is among
the brightest 2\% of the 811 objects in the total LBG
spectroscopic sample presented in \citet{shapley2003}; C49 would have been 
among the brightest 8\%. 

\begin{figure*}[t!]
\epsscale{0.7}
\plotone{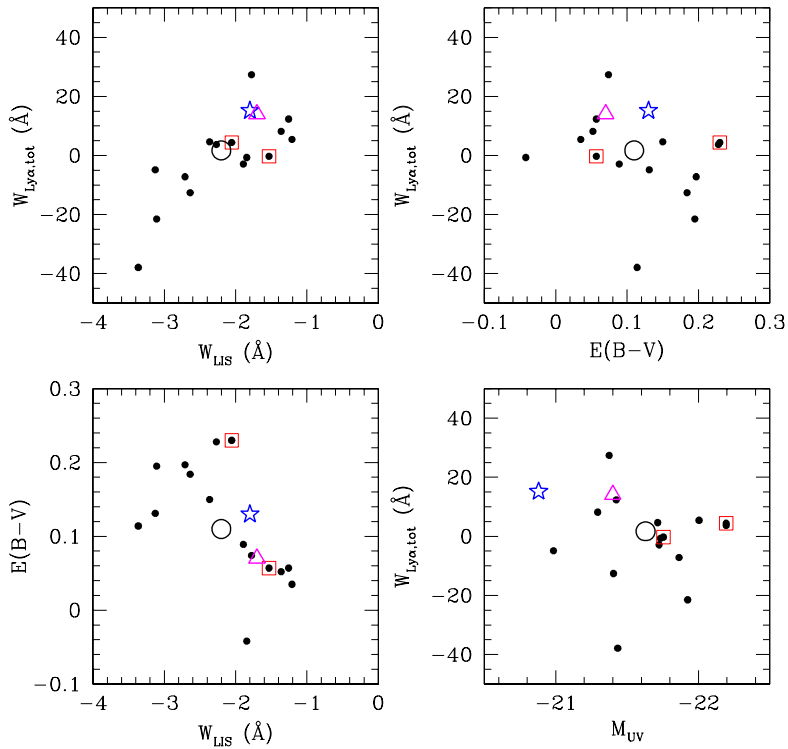}
\epsscale{1.0}
\caption{Rest-frame UV properties
of the deep spectroscopic sample. In each panel, solid black points
indicate individual objects in the deep spectroscopic sample;
points for C49 and D3 are enclosed in open squares; the large open
circle, star, and triangle indicate the average values for
the deep spectroscopic sample, the total LBG spectroscopic
sample from \citet{shapley2003}, and the original
Lyman-continuum sample of \citet{steidel2001}, respectively.
In general, while the \citet{steidel2001}
sample is very similar to the total LBG sample in terms of average
Ly$\alpha$ and low-ionization interstellar equivalent widths,
the current sample is much closer to the total LBG sample in terms
of average UV extinction properties.
(Upper left) The correlation between total Ly$\alpha$ equivalent
width and low-ionization interstellar absorption equivalent width
in composite spectra \citep{shapley2003}
is confirmed on an object-by-object basis for the deep sample.
Positive values of equivalent width indicate emission.
(Upper right) The correlation between total Ly$\alpha$ equivalent
width and UV extinction.
D3 has the largest inferred value of $E(B-V)$ in the deep sample.
(Lower left) The correlation
between $E(B-V)$ and low-ionization
interstellar absorption equivalent width. In this plot,
D3 stands out for having an anomalously high dust extinction,
given the equivalent width of its low-ionization interstellar
absorption lines. (Lower right) A plot of total Ly$\alpha$
equivalent width as a function of UV absolute magnitude,
uncorrected for dust extinction. D3 is the brightest object in the
sample. Also, it is clear that both the current and \citet{steidel2001}
samples are significantly more luminous on average than the total LBG
spectroscopic sample.
}
\label{fig:specpanel}
\end{figure*}

We have interpreted the rest-frame UV colors of galaxies in the
current deep sample assuming that
the unreddened continuum is well-represented
by a 300~Myr continuous star-formation model, where 300~Myr
represents the median age in the LBG sample analyzed
by \citet{shapley2001}.  Differences
in $G-{\cal R}$ color (corrected for IGM absorption and Ly$\alpha$
emission/absorption intrinsic to the galaxies) are interpreted
as different amounts of dust extinction, parameterized by
a \citet{calzetti2000} attenuation curve.
Accordingly, the observed
$G-{\cal R}$ colors can be converted into $E(B-V)$ values,
with the largest associated uncertainties resulting from the uncertainties in
assumed underlying stellar population template, and the observational
errors on $G-{\cal R}$ color (typically at the $5-10$\% level).
We note that estimates of UV-continuum slope and $E(B-V)$ inferred directly
from the shapes of the deep spectra redwards of Ly$\alpha$
are not very robust, because the changes in relative
flux level corresponding to different degrees of UV extinction are
not significant over the range covered by the spectra.

As shown in Figure~\ref{fig:specpanel}, D3 has $E(B-V)=0.23$, which implies
the highest level of extinction in the deep sample, more
than 2 magnitudes of attenuation at rest-frame UV wavelengths. 
D3 is significantly
redder than the average in the deep sample, the total LBG
sample from \citet{shapley2003}, and the sample of LBGs
used by \citet{steidel2001} to construct a composite spectrum
that showed significant Lyman-continuum emission (see 
Table~2). In fact, the level of dust extinction in D3 may be even 
higher; simultaneously
fitting the $G-{\cal R}$ and ${\cal R}-K_s$ colors of D3 \citep{shapley2001} 
yields an age $\leq 10$~Myr, and $E(B-V)\geq 0.3$ (i.e. the assumption
of 300~Myr for the age will lead to an underestimate of the extinction
if the true age is younger). 
On the other hand, \citet{reddy2006} find that LBGs with the 
youngest inferred ages do not appear to follow the \citet{meurer1999}
relation between UV-reddening and dust obscuration, in the
sense that they have redder UV colors for a given amount
dust extinction. The young apparent age of D3 might therefore
cause us to overestimate the true level of dust obscuration.
As we return to in section~\ref{sec:fescape}, 
the heavy dust extinction inferred
from the UV colors of this object is somewhat difficult to reconcile
with the significant Lyman-continuum emission.
In fact, \citet{steidel2001} propose that it is galaxies
with less reddened continua that are more likely to show escaping
Lyman-continuum emission. We have also checked whether the composite
nature of D3 could have led to a bias in the measured $G-{\cal R}$
color, but find that the morphology between the $G$ and ${\cal R}$
bands remains roughly constant (which is consistent with
the roughly constant ratio between the continua of D3-ap1 and D3-ap2
spectra spanning from 4000~\AA\ to 6300~\AA). Therefore, the ${\cal R}$-band
detection isophote applied to the $G$ band should yield an
accurate representation of the $G-{\cal R}$ color. In contrast to
D3, the other object with significant Lyman-continuum emission,
C49, has an inferred $E(B-V)=0.06$. This low value of
$E(B-V)$ places C49 among the third of the deep sample with
the least extinction. It appears significantly less reddened
than the average of the deep sample and the total LBG sample of 
\citet{shapley2003}, and even slightly bluer than the average
of the blue sample presented in \citet{steidel2001}
(see Figure~\ref{fig:specpanel}). Thus, no 
consistent picture emerges from D3 and C49 
about the dust extinction properties of galaxies with significant
escaping Lyman-continuum emission.

Now we compare the rest-frame UV spectroscopic features of D3 and C49
with those of the samples listed in Table~2.
Ly$\alpha$ is observed in emission for both components
of D3, with a rest-frame equivalent width of $\sim 5$~\AA\ in
both apertures. C49 also shows weak Ly$\alpha$ emission, but
superposed on an absorption trough of roughly equal strength,
such that the total equivalent width is $\sim 0$\AA. 
The Ly$\alpha$ emission in C49 is actually double-peaked, though this
fact is not readily apparent in Figure~\ref{fig:lycdet}, 
which has been smoothed by a 3-pixel boxcar. 
As shown in Figure~\ref{fig:specpanel},
D3 and C49 exhibit close to the average Ly$\alpha$
equivalent width in the deep sample, which is itself characterized
by weaker Ly$\alpha$ emission on average than the larger LBG sample
of \cite{shapley2003} and the sample used to detect Lyman-continuum 
emission in \citet{steidel2001}.

Another relevant property is the strength of the low-ionization 
interstellar metal absorption lines. 
The strongest low-ionization interstellar absorption lines in the
spectrum of D3 (aperture 1) are consistent with zero intensity at line center;
in this aspect, the D3 spectrum may
imply a unity covering fraction of gas giving rise to
low-ionization metal absorption, at least at
some velocities. As shown in Figure~\ref{fig:specpanel},
the low-ionization absorption equivalent widths in D3 
are close to the mean of the deep sample, which is stronger
than the average in both the \citet{shapley2003} and
\citet{steidel2001} samples; D3 also has narrower low-ionization
velocity widths than the deep sample average, as estimated
from the Si~II $\lambda 1260$ and C~II $\lambda 1334$ absorption lines. 
Also shown in Figure~\ref{fig:specpanel},
C49 has weaker than average low-ionization
absorption lines that do not reach zero intensity at line
center, even after deconvolution of the spectral point spread function.
In the absence of unresolved saturated components for which
we cannot correct, it appears that C49 is characterized
by neutral gas with non-unity covering fraction -- i.e. a patchy
spatial distribution. The strength of the low-ionization
interstellar absorption lines in the spectrum of C49
is very similar to those measured from
the composite spectrum of \citet{steidel2001}, in which significant
Lyman-continuum emission was detected. Weak low-ionization interstellar
absorption does not however appear to be a sufficient condition for
escaping Lyman-continuum emission. Three
galaxies in the deep sample (C32, D17, and MD23)
have weaker low-ionization absorption 
lines than C49, and  C32 and D17 also appear to have
lower covering fractions
of neutral gas. However, 
none of these three galaxies exhibit any significant emission in the 
Lyman-continuum region. 

One striking property of D3 is the apparent
weakness of Lyman series absorption. The noise in the spectrum
at these wavelengths prevents a precise determination of 
H~I column density, yet a comparison with the average spectrum
of the deep sample reveals significantly weaker than average 
absorption in D3 from Ly$\beta$, Ly$\gamma$, Ly$\epsilon$. 
In particular, the component
of D3 with detected Lyman-continuum emission is quite
distinct from the rest of the deep sample in terms of the weakness
of Ly$\beta$ absorption. On the other hand, C49
agrees very well with the deep sample average for
Ly$\beta$ and the higher-order Lyman series absorption features.

The spectra of D3 and C49 feature both Ly$\alpha$
emission and interstellar absorption lines, and the higher redshift
of Ly$\alpha$ emission relative to that of the
absorption lines indicates kinematic
evidence for large-scale outflow motions. This kinematic signature 
appears to be an almost universal characteristic of
LBG rest-frame UV spectra \citep{pettini2001,shapley2003}. 
The velocity offset between Ly$\alpha$ emission and interstellar
absorption lines is $\sim 500$~\kms\ in D3 (aperture 1)
and somewhere between $780$ and $1130$~\kms\ in C49, where the
range for C49 indicates the different offsets calculated using each of the two
peaks of Ly$\alpha$ emission. The velocity offset in D3 is
smaller than the deep sample average for objects with
both emission and absorption, as well as the average
offset measured by \citet{shapley2003}. C49,
on the other hand, shows a larger than average offset between
emission and absorption velocities. Additional rest-frame optical
nebular emission line measurements of D3 \citep{pettini2001} indicate
that the low-ionization interstellar absorption lines are
blueshifted by $\sim 280$~\kms\ relative to the H~II regions,
while Ly$\alpha$ is redshifted by only $\sim 215~$\kms. Therefore,
in D3, it is the redshift of Ly$\alpha$ relative to H~II regions
which is smaller than the typical value found in LBGs, not
the blueshift of the low-ionization interstellar absorption lines,
and the associated outflow speed \citep{pettini2001,steidel2004}.
It will be valuable to obtain H~II region emission line measurements
for C49 as well, in order to locate the systemic redshift
for this system, and determine the outflow speed from the 
blueshift of interstellar absorption lines.

Although the higher-ionization
interstellar absorption lines are noisy,
we can roughly estimate that the Si~IV
$\lambda \lambda 1393,1402$ absorption is comparable in D3
to that in the average spectrum of the deep sample. The Si~IV
profiles are too noisy to determine a useful estimate of 
the velocity widths. It is not possible to measure
the strength or width of the C~IV line in D3, as the feature
is contaminated by a strong sky residual. In C49, both Si~IV
and C~IV interstellar absorption are weaker than the average of
the deep sample. Additionally, the Si~IV lines are significantly
narrower than the average Si~IV velocity width in the deep sample.

\subsubsection{Additional Properties of D3}
\label{sec:resultD3C49d3}

\begin{figure*}[t!]
\plottwo{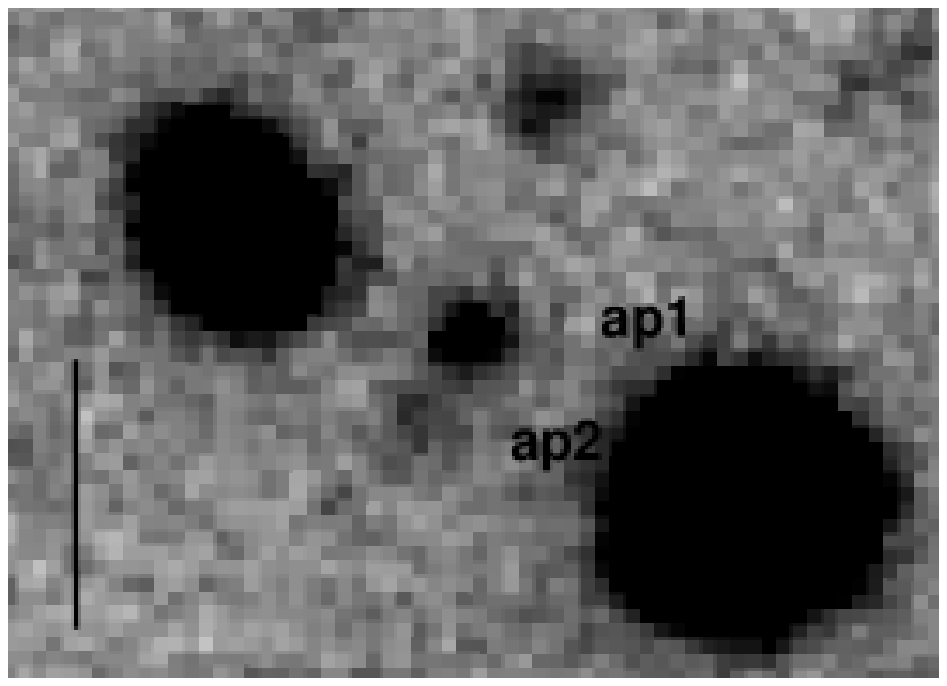}{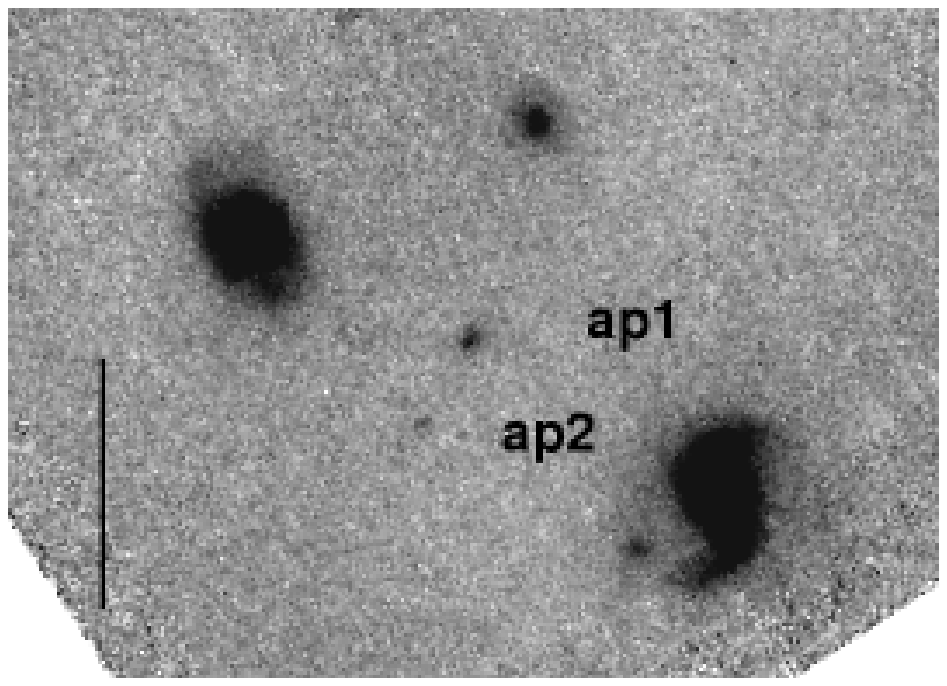}
\caption{${\cal R}$ (left) and {\it HST}/NICMOS F160W (right)
images of D3. The ${\cal R}$ image was obtained at
the Palomar 200 inch Hale Telescope, under $\sim 1$" seeing
conditions, while the F160W image was obtained with the
NIC2 camera of NICMOS, which provides diffraction-limited
(0\secpoint14 resolution) imaging at $1.6\mu$m.
The locations of spectral apertures 1 and 2 are marked
in each image, as ``ap1" and ``ap 2," respectively.
The vertical bar in the lower left-hand
corner of each figure indicates a scale of 5". D3-ap1
is the component for which flux was detected in
the Lyman continuum region. Both the
${\cal R}$ and {\it HST}/NICMOS F160W
images indicate the brighter and more compact nature
of D3-ap1, relative to D3-ap2.
D3 is the only object in the deep sample with confirmed
double morphology.
}
\label{fig:D3img}
\end{figure*}

D3 is one of the brightest objects in the total spectroscopic
LBG sample, and, as such, has served as an obvious target for multi-wavelength
follow-up. The combination of Keck/NIRC
$K_s$-band photometry with existing optical observations 
indicates a young age for D3, of less than 10~Myr, and a stellar mass of
$\sim 6\times 10^{9} M_{\odot}$, assuming a Chabrier IMF \citep{shapley2001}. 
The intrinsic ratio of Lyman-continuum to non-ionizing UV luminosity
is a decreasing function of age, so that the highest intrinsic ratio
would be found in galaxies with ages as young as that inferred for D3 --
though, of course, the true age for D3 cannot be less than
10~Myr, i.e., shorter than its dynamical timescale.
Observations of C49 in the near-IR constitute a natural
follow-up program and will allow similar
age and stellar mass determinations.
Keck/NIRSPEC Near-IR spectroscopy covering the 
[OIII]~$\lambda\lambda 5007,4959$
H$\beta$, and [OII]~$\lambda 3727$ emission lines enables
an estimate of the $R_{23}$ chemical abundance
indicator \citep{pagel1979}. Based on the value
of $R_{23}$, \citet{pettini2001} find O/H=$0.24-0.78$ (O/H)$_{\odot}$
for D3, where these values refer to the revised solar oxygen
abundance of $12+\log(\mbox{O/H})_{\odot} = 8.66$ \citep{asplund2004}. 
The NIRSPEC spectrum is roughly centered on the
D3 component showing significant Lyman-continuum emission,
and the slit was not positioned at an angle to cover
both components. Further near-IR information includes
three orbits of {\it HST}/NICMOS F160W diffraction-limited imaging.
As shown in Figure~\ref{fig:D3img},
the rest-frame optical image confirms with significantly
sharper resolution what is suggested in the ${\cal R}$ band
image of D3. Specifically, the component with escaping 
Lyman-continuum emission has a peak surface brightness $2.5-3.0$ times
higher than that of the other component. The
brighter component is also more compact and radially-symmetric, 
with a diameter roughly half the extent of the longest linear dimension
of the extended, irregular and diffuse emission of the second
component. Longer wavelength observations of
D3 thus indicate the complex morphology of a young
system with a warm ionized gas phase that has already been enriched
to a significant fraction of solar metallicity.

As the redshift of D3 places it within the well-studied overdensity 
of LBGs at $z= 3.09\pm 0.03$ \citep{steidel1998,steidel2000},
narrow-band imaging can be used to reveal the spatial
distribution of Ly$\alpha$ emission in this object. The narrow-band
observations of \citet{matsuda2004}
indicate that Ly$\alpha$ emission is extended
over 17 arcmin$^2$ with a luminosity of 
$9.5 \times 10^{42}\mbox{erg}\mbox{ s}^{-1}$, 
and roughly centered on the ${\cal R}$-band
position of the component with detected Lyman-continuum emission.
The large spatial extent of its Ly$\alpha$ emission places D3 in the
so-called Ly$\alpha$ ``blob'' sample of \citet{matsuda2004}.
The Ly$\alpha$ luminosity reflects the number ionizing photons
absorbed, and re-emitted as Ly$\alpha$ photons by recombining
gas. For more than one third of the ``blob''
sample, the number of intrinsic Lyman-continuum photons inferred
from the UV continuum is exceeded by the number
inferred from the Ly$\alpha$ luminosity. Potential
explanations for the excess of Ly$\alpha$ emission
include photoionization by a young, metal-poor stellar population 
with a flatter than Salpeter IMF at the high-mass end; 
by an AGN; by hidden ionizing
UV sources; or by the diffuse intergalactic background radiation.
Figure 10 of \citet{matsuda2004}, showing the
star-formation rate inferred from Ly$\alpha$
compared with that inferred from the non-ionizing
UV-continuum, indicates that no such explanation is
required for D3. The number of ionizing photons
reprocessed into Ly$\alpha$ emission represents
only $\leq 25$\% of the number inferred from the 
non-ionizing UV continuum. Understanding the
extended Ly$\alpha$ emission in D3 is yet
another goal in untangling the complex nature
of this object. Since the redshift of C49 does
not place it within the SSA22a galaxy overdensity,
no information is presently available about the
spatial distribution of its Ly$\alpha$ emission.

In summary, no consistent picture emerges from D3 and C49
about the properties of galaxies with escaping
Lyman-continuum emission. While C49 is one the 
bluest galaxies in the deep sample,
D3 has the reddest observed UV continuum.
While the low-ionization interstellar metal absorption lines
from C49 indicate a non-unity covering fraction of
cool gas, favorable for allowing the escape of Lyman-continuum
photons, there are galaxies in the deep sample
with weaker low-ionization interstellar absorption
lines and also apparently lower covering fractions
of cool gas, which at the same time do not
exhibit significant escaping ionizing radiation. Furthermore,
D3, which does exhibit escaping Lyman-continuum
radiation, has low-ionization interstellar absorption lines
that may be black at line center, indicating a unity covering fraction
of cool gas. Two striking properties of D3 are its complex double
morphology, consisting of a compact component exhibiting
Lyman-continuum, and a fainter and more diffuse component with 
no corresponding detection; and its weak Lyman series absorption --
significantly weaker than the typical level of absorption in the
deep sample. At the same time, C49 has a simple single-component
morphology, and Lyman series absorption that agrees quite well
with that in the sample average spectrum. Clearly, a sample
of two galaxies is too small
to infer the typical properties of galaxies with escaping
Lyman-continuum emission. 

\section{The Escape Fraction of Ionizing Radiation}
\label{sec:fescape}
The detection of escaping Lyman-continuum emission from individual
star-forming galaxies at $z\sim 3$ 
represents definite forward progress from previous
statistical detections \citep{steidel2001} and individual
non-detections \citep{giallongo2002,inoue2005,dawson2002}. 
The range of detections and non-detections of ionizing
flux in our sample can also be combined with measured
1500~\AA\ fluxes to characterize the emergent far-UV
spectral shapes of star-forming galaxies at $z\sim 3$.
Ultimately, our goal is to quantify the global contribution
of LBGs to the ionizing background at $z\sim 3$. The most realistic
route for addressing this problem is to convert the well-characterized
LBG 1500~\AA\ luminosity function \citep{steidel1999,adelberger2000}
into a Lyman-continuum luminosity function using
the average emergent ratio of 1500~\AA\ to Lyman-continuum
flux densities, corrected for IGM absorption, and determined from 
a small but representative sample of LBGs.

From a theoretical perspective \citep{loeb2001,furlanetto2004,cen1993}, 
it is of interest to determine
the absolute escape fraction of Lyman-continuum photons, $f_{esc}$. This
quantity, defined as the ratio of $900$~\AA\ photons produced that
escape from the galaxy without being absorbed, relative to the
total number produced, is difficult to estimate from even the most
sensitive observations, because it relies on knowledge of
the intrinsic number of Lyman-continuum photons
produced. A related quantity is $f_{esc, rel}$,
first defined by \citet{steidel2001} as the fraction
of escaping Lyman-continuum photons normalized by the
fraction of escaping 1500~\AA\ photons. 
In practice, $f_{esc, rel}$ is a
better defined observational quantity, since it can be combined with the 
well-determined UV continuum luminosity distribution to estimate 
the global contribution of LBGs to the ionizing background. It is also
possible to determine $f_{esc}$ from $f_{esc, rel}$ if the degree of
dust extinction in the UV is known. 
The relative escape fraction, $f_{esc, rel}$,
can be expressed in terms of the observed 1500~\AA\ to 
900~\AA\ flux-density ratio, $f_{1500}/f_{900}$, as:

\begin{equation}
f_{esc, rel} = \frac{(L_{1500}/L_{900})_{int}}{(f_{1500}/f_{900})}\exp(\tau_{IGM,900})
\end{equation}

where $(L_{1500}/L_{900})_{int}$ is the intrinsic ratio of
non-ionizing to ionizing specific intensities, and 
$\tau_{IGM,900}$ represents the line-of-sight
opacity of the IGM to Lyman-continuum photons, which
can be estimated either
empirically \citep{steidel2001} or through
simulations \citep{madau1995,bershady1999,inoue2005}.
The real difficulty in estimating $f_{esc,rel}$ and $f_{esc}$
stems from the fact that the intrinsic spectral
shape for massive stars above and below
the Lyman Break is not well-constrained by observations.
$(L_{1500}/L_{900})_{int}$  must therefore be estimated from stellar 
population synthesis models, and varies in the models as a
function of stellar population age, metallicity,
star-formation history, and IMF \citep{bc2003,leitherer1999}.
Even when fixing the IMF and star-formation history, \citet{inoue2005} 
find that $(L_{1500}/L_{900})_{int}$ varies from $1.5-5.5$,
mainly as a function of time since the onset of 
continuous star-formation 
(where the units of $L_{1500}$ and $L_{900}$ are 
$\mbox{ erg s}^{-1}\mbox{Hz}^{-1}$).
Both \citet{steidel2001} and \citet{inoue2005}
adopt an intrinsic ratio of $(L_{1500}/L_{900})_{int}=3.0$. For
comparison with previous work, we adopt the same ratio
here, but note the significant associated uncertainties.

\begin{figure}
\plotone{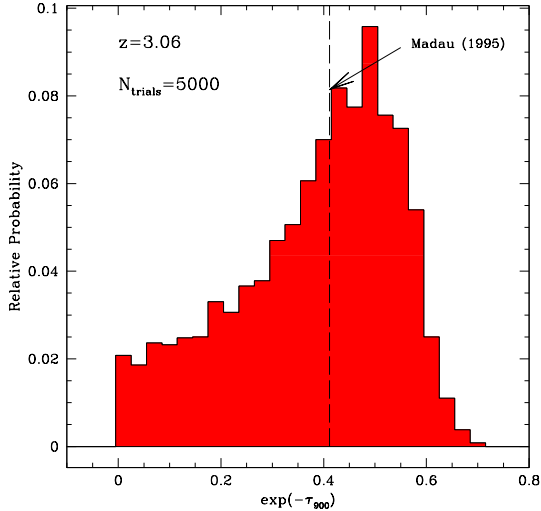}
\caption{Results of Monte Carlo simulations of IGM opacity
along the line of sight to $z=3.06$, the average redshift of
the deep sample. A large number $N=5000$ of random
sightlines to $z=3.06$ were generated. For each sightline,
the ratio of the attenuated to unattenuated
flux densities in the rest-frame $880-910$~\AA\ was computed,
and is expressed here as $\exp(-\tau_{900})$. The attenuation
factor discussed in the text consists of the inverse
of $\exp(-\tau_{900})$. At $z=3.06$, we find
$\langle \exp(-\tau_{900}) \rangle= 0.38$, corresponding
to an attenuation factor of $2.6$. The 95.4\% confidence
interval for the $880-910$~\AA\ attenuation factor at $z=3.06$
spans from 35 to 1.6. The dotted line indicates
the analytic approximation of \citet{madau1995} for the
mean transmission in the $880-910$~\AA\ region at
$z=3.06$. The tail in the distribution extending down to zero transmission
is dominated by sightlines in which at least one Lyman
limit system fell within the redshift range corresponding
to rest-frame $880-910$~\AA.
}
\label{fig:tauhist}
\end{figure}

In this section, we present estimates of $f_{1500}/f_{900}$ 
and $f_{esc, rel}$ for the two individual detections, 
as well as an estimate of the sample average $f_{1500}/f_{900}$
and $f_{esc,rel}$.  Modeling and correcting for the 
effects of IGM Lyman-continuum
opacity enables the conversion of a raw $f_{1500}/f_{900}$ value
to the emergent $f_{1500}/f_{900}$$,_{corr}$ that would
be observed in the immediate vicinity of a galaxy, which is one of
the quantities required for estimating the global
contribution of LBGs to the ionizing background at $z\sim 3$.
In order to model IGM opacity,
we use Monte Carlo simulations of intergalactic absorption
to generate large samples of random sightlines 
to the redshifts of C49 and D3,
and to the average redshift of the deep sample. Absorbers
are drawn at random from the column-density and redshift
distributions constituting the $MC-NH$ model of \citep{bershady1999},
and then applied to the unabsorbed continuum of each simulated
sightline. The resulting sample of model sightlines is
analyzed to obtain the mean and standard deviation of
the optical depth in the rest-frame region $880-910$~\AA\
at each emission redshift. As an example, we show
the simulated distribution of optical depth for the 
sample average redshift, $\langle z \rangle = 3.06$, 
in Figure~\ref{fig:tauhist}. In order to correct the composite
LBG spectrum in \citet{steidel2001}, a more
empirical approach to estimate the IGM opacity at $z=3.47$ was adopted.
A sample of 15 bright QSO spectra at $\langle z\rangle =3.47$
were combined to create a composite spectrum with
no intrinsic Lyman-continuum opacity, yielding 
the average intervening absorption in the Lyman continuum
over 15 random sightlines. The absorption estimated
in this empirical fashion was a factor of $\sim 3.8$,
which agrees well with the estimate of a factor of $\sim 3.7$
from our simulated $z=3.47$ sightlines.
Finally, in this section, we also attempt to quantify the sample variation in 
$f_{1500}/f_{900}$$,_{corr}$, taking into account sightline to
sightline variations in IGM opacity.

\subsection{The Emergent UV Spectrum: C49, D3, and the Sample Average}
\label{sec:fescapeobs}
The observed UV-to-Lyman-continuum flux-density ratio in C49 is
$f_{1500}/f_{900} = 12.7\pm 1.8$. Simulations of IGM
neutral hydrogen opacity indicate that,
on average, the flux level in the rest-frame range $880-910$~\AA\ 
is attenuated by a factor of $2.8$
for sources at $z=3.15$ (the redshift of C49). If the
line of sight to C49 is typical for its redshift,
then the emergent far-UV flux-density ratio from the galaxy
is $f_{1500}/f_{900}$$,_{corr} = 4.5$, very similar to that 
inferred from the composite LBG spectrum \citet{steidel2001}.
Furthermore, the rest-frame UV colors of C49 imply that differential
dust extinction between the UV and Lyman-continuum is not
significant. Assuming the same intrinsic ratio of $(L_{1500}/L_{900})_{int}$
as \citet{steidel2001}, we find $f_{esc,rel}=65$\%
for C49. If C49 is, however, an unusually transparent sightline,
the emergent UV-to-Lyman-continuum ratio is larger. Our
Monte Carlo simulations of IGM opacity indicate that, 
at a 95.4\% (i.e. 2$\sigma$) confidence level,
the attenuation in the $880-910$~\AA\
rest frame falls between a factor of 38 and 1.7.
At the transparent edge of the 2$\sigma$ confidence interval,
the IGM attenuation implies 
$f_{1500}/f_{900}$$,_{corr} = 7.5$, and $f_{esc,rel}\sim 40$\%.
Given that continuum flux is detected in C49 down to $\sim 800$\AA,
it is likely for the transparency to be higher than average in the
$\sim 880-910$~\AA\ range.
It is important to note here that one cannot infer
the transparency level in the range $\sim 880-910$~\AA,
based on $D_A$, the continuum decrement at $1050-1170$~\AA.
$D_A$ indicates absorption from intervening Ly$\alpha$ lines,
and is dominated by low-column-density
systems (i.e. $N(HI) < 10^{15}\mbox{ cm}^{-2}$).
On the other hand, the
Lyman-continuum attenuation factor is modulated by
the number of intervening systems with $N(HI) > 10^{17}\mbox{ cm}^{-2}$
at redshifts such that the associated Lyman-limit absorption 
falls in the window of interest.
Accordingly, our Monte Carlo IGM simulations
indicate the absence of a correlation between $D_A$ and
the attenuation factor at $\sim 880-910$~\AA.

In D3, the UV-to-Lyman-continuum flux-density ratio is estimated
using the $1500$~\AA\ flux from both components, compared
with the Lyman-continuum detection from aperture~1, alone.
This comparison yields $f_{1500}/f_{900} = 7.5 \pm 1.0$. Correcting for
the average $880-910$~\AA\ attenuation of a factor
of $2.6$ at $z=3.07$ (the redshift of D3), we
find $f_{1500}/f_{900}$$,_{corr} = 2.9$, which is roughly the
same as the intrinsic ratio of $L_{1500}/L_{900}$ assumed
by \citet{steidel2001} and \citet{inoue2005}, and
therefore implies $f_{esc,rel}\geq 100$\%! 
Again, it is worth noting that
the inferred significant dust extinction in D3
is difficult to reconcile with such an extreme
value of $f_{esc,rel}$, as a reasonable extrapolation
of the \citet{calzetti2000} attenuation law would
imply significantly more attenuation at 900~\AA\
than at 1500~\AA.

There are multiple ways to explain the extreme and apparently
contradictory properties of D3. First, in addition 
to the possibility that interpreting the UV color with the Calzetti 
law causes us to overestimate the overall dust extinction in D3 
(see section~\ref{sec:resultD3C49uv}), 
the $E(B-V)$ inferred for this object may only 
suffice as a global quantity, averaged over the entire
surface area of the galaxy. If, in fact, the escaping Lyman-continuum
emission is patchy, and preferentially escaping from regions
with less dust extinction, it may not be appropriate to use
the {\it average} $E(B-V)$ to interpret the emergent ratio
of UV to Lyman-continuum radiation. Future {\it HST}/ACS
imaging in two filters may constrain the detailed dust extinction 
distribution in this source. Second, 
the average correction for IGM opacity
along the line of sight may not apply to an individual
sightline. For a given emission redshift, there is
significant variation in the rest-frame $880-910$~\AA\
opacity from sightline to sightline. In our Monte Carlo simulations
of IGM opacity, we found that, at 95.4\% (2$\sigma$) confidence,
the attenuation in the Lyman continuum at $z=3.07$
ranged between 35 and 1.6. Using 1.6 as a 
lower limit on the IGM attenuation, we derive
$f_{1500}/f_{900}$$,_{corr} = 4.7$. Compared to an assumed
$(L_{1500}/L_{900})_{int}$ ratio of 3, this yields
$f_{esc,rel}\sim 65$\%.
As in the case of C49, adopting a lower than average IGM opacity in the Lyman
continuum may indeed be suggested by the persistence
of the D3 continuum down to rest wavelengths of $\sim 810$~\AA,
well below one Lyman-continuum mean free path.
Third, we consider the
possibility that the intrinsic UV continuum of D3 is
not dominated by stars, but by the emission from an
AGN, with a smaller intrinsic ratio of $(L_{1500}/L_{900})_{int}$.
This scenario is not supported by rest-frame UV spectroscopic
observations. We detect no strong C~IV or N~V emission in the
rest-frame UV spectrum, which would be characteristic. Furthermore,
{\it HST}/NICMOS high-resolution imaging reveals no unresolved
nuclear emission in the rest-frame optical. Future X-ray and
mid-IR observations with {\it Chandra} and {\it Spitzer},
respectively, may constrain the contribution of an AGN to the emission
in D3. 

Finally, there is the possibility that the apparent detection
of flux below the Lyman limit in D3 results from scattered light from 
the outer isophotes one or both of the bright galaxies within several arcseconds
of the position of D3-ap1. These bright neighbors stand out
in the $\cal{R}$ and F160W images in
Figure~\ref{fig:D3img}, with centroids 5\secpoint4 to the northeast of 
D3-ap1, and 5\secpoint9 to the southwest. Perhaps most relevant,
the brighter southwest object
has $U_n=21.6$, while the northeast object has $U_n=23.5$.
D3 is unique among the objects in the
deep sample, in terms of being flanked by such bright objects.
While the mask position angle was chosen to be roughly parallactic
for exposures at hour angles east of overhead, the direction of differential
refraction for exposures obtained when the SSA22a field was
setting is such that light from the brighter southwest object was
bent in the direction of the D3 slit. However, even in the
worst case scenario, the deflection at
3600~\AA\ for the setting exposures with the maximum 
airmass is still only 0\secpoint64, a small fraction of the distance
between the bright southwest object and the edge of the D3 slit. 
Nonetheless, the unusual configuration of D3 with its bright neighbors
should be noted.

We also consider the mean observed UV-to-Lyman-continuum flux-density ratio for
the entire sample of 14 objects with deep spectra, by constructing
a composite spectrum for which each individual galaxy spectrum was scaled
to a common mode at $1400-1500$~\AA\ before averaging. This mean ratio
is $\langle f_{1500}/f_{900} \rangle = 58\pm 18_{stat} \pm 17_{sys}$, where the
two sources of error reflect the pixel to pixel noise in the
composite spectrum and the systematic uncertainties in the individual values
of $f_{900}$ due to flatfielding and sky-subtraction, respectively.
The mean redshift of the sample is $\langle z \rangle =3.06$, 
very close to that of D3. Therefore, the same correction
factor for IGM opacity (2.6) can be applied to this ratio to obtain
the average emergent UV-to-Lyman-continuum ratio. With
this correction, we obtain $\langle f_{1500}/f_{900}$$,_{corr}\rangle = 22$.
Assuming an intrinsic UV-to-Lyman continuum ratio of 
$(L_{1500}/L_{900})_{int}=3.0$, as in \citet{steidel2001},
we find $\langle f_{esc,rel} \rangle =14$\% 
for the deep sample --  a factor of $\sim 4.5$ lower than 
the value of $f_{esc,rel}$ reported in \citet{steidel2001}.
We will return to this result in section~\ref{sec:implications}.
While significantly smaller than the uncertainty in opacity for an individual
sightline, the average opacity for a sample of 14 sightlines
has an associated error, because of the finite number of sightlines
being averaged together. To estimate this uncertainty,
we drew a large number of random samples of 14 sightlines from our 
simulations of single IGM sightline opacities. For each random sample,
we computed the average IGM opacity, and then computed the mean
and width of the distribution of IGM opacities averaged
over 14 sightlines. With 95.4\% ($2\sigma$) confidence,  we found
that the average IGM opacity for the sample of 14 sightlines requires
a correction factor between $2.2$ and $3.3$, implying that
$\langle f_{1500}/f_{900}$$,_{corr}\rangle$ falls between 
$17$ and $27$. The corresponding range in $f_{esc,rel}$
is 11\%$-$18\%, assuming $(L_{1500}/L_{900})_{int}$ as before.

There are two important caveats associated with our attempts
to model and correct for IGM opacity and infer the shape
of the emergent far-UV spectrum, $f_{1500}/f_{900}$$,_{corr}$.
First, the IGM simulations we performed to characterize
the distribution of attenuation factors in the Lyman-continuum
region are based on the assumption that each random sightline is
a probing the average properties of the IGM, and do
not include the effects of large-scale structure.  In fact,
nine of 14 of the galaxies in our sample have redshifts
that place them within the significant overdensity
in the SSA22a field
discovered by \citet{steidel1998}, which is characterized by
$\delta_{gal}\sim 6$ at $\langle z \rangle = 3.09 \pm 0.03$.
Most likely, the space density of optically-thick 
Lyman limit systems with $N(HI)> 10^{17}\mbox{cm}^{-2}$
is enhanced within the redshift range of the protocluster,
tracking the galaxy density. Our simulations indicate
that the attenuation in the $880-910$~\AA\ range is more than
twice as high for sightlines with intervening Lyman limit systems
within one Lyman-continuum mean free path
of the emission redshift, as opposed to sightlines without
such absorbers. Therefore, we expect a corresponding enhancement
in the average rest-frame $880-910$~\AA\ attenuation for
a composite spectrum dominated by protocluster galaxies.
The magnitude of the enhancement in the average attenuation
is not obvious, however, as demonstrated by the cases of
our two individual Lyman-continuum detections.
The sightline of C49 ($z=3.15$) passes through
the overdensity, for example, yet it features very little if any 
Lyman-continuum absorption at $3.06 \leq z \leq 3.12$. The redshift of
D3 $(z=3.07)$ places it within the protocluster, yet its spectrum also
appears fairly transparent in the Lyman-continuum region.
A more careful treatment accounting for the relative spatial
distributions of absorbers and galaxies is required to quantify the effect
of the protocluster on the average IGM opacity in the Lyman-continuum
region for our sample, yet we can state qualitatively that
the $z=3.06$ average Lyman-continuum attenuation factor of 2.6
likely represents an underestimate.

The second, related, caveat stems from the assumption in the
IGM simulations that each randomly-generated sightline is probing a volume
that is statistically-independent from the volumes probed
by all other random sightlines.
Since our sample is contained within an area of $9\times13$ comoving
Mpc$^2$ on the sky, $\sim 50$\% of the resulting set of galaxy sightline pairs
are separated by less than 5.7 comoving Mpc, the auto-correlation length
of $z\sim 3$ LBGs \citep{adelberger2005a}. This auto-correlation
length is relevant since it has been demonstrated
that, at least at $N(HI)>1\times 10^{17}\mbox{cm}^2$, intervening
absorbers are associated with galaxies and 
therefore have similar clustering properties 
\citep{steidel1990,adelberger2003}.
The IGM opacity along neighboring sightlines in our sample
must then be correlated, so it is not strictly correct to 
treat the galaxy sightlines as
statistically independent of each other during our simulations.
The above caveats should be borne in mind throughout the
remainder of the discussion.

\subsection{Variation in The Emergent UV Spectrum}
\label{sec:fescapevar}
It is very striking that C49 and D3 appear to have such significant
Lyman-continuum fluxes and low ratios of 
UV to Lyman-continuum flux densities, while the remainder of objects in
the sample are characterized by $f_{900}$ non-detections. 
If there is significant variance
in the emergent far-UV spectral shape in LBGs, understanding the
origin of the variance will be a crucial component of characterizing
the reionization of the universe. Therefore, the apparent variance
in $f_{1500}/f_{900}$ must be quantified more concretely, and
related to the range of emergent far-UV spectral shapes, corrected
for IGM absorption (i.e. $f_{1500}/f_{900}$$,_{corr}$).

First, as the sample spans a range in $f_{1500}$, it is important
to confirm the fraction of our sample for which 
the observed $f_{1500}/f_{900}$ ratios of C49 and D3 are detectable
at the $3\sigma$ level.
Since C49, for example, has a $1500$~\AA\ flux density that is above
the median in the sample, it is possible to
detect larger $f_{1500}/f_{900}$ ratios in this object than
in an object such as MD14, which has one of the faintest $1500$~\AA\
flux densities in the sample. By comparing the $1500$~\AA\ fluxes
of all the objects, with the $3\sigma$ pixel to pixel noise limits in the
$880-910$~\AA\ wavelength range, we find that the observed
$f_{1500}/f_{900}=7.5$ for D3 could have been detected in
all of the objects. The UV-to-Lyman-continuum flux-density ratio
in C49, $f_{1500}/f_{900}=12.7$, could have been detected
in all but the two faintest objects, D7 and MD14. In fact, 
we find that the median detectable flux-density ratio for individual
objects in the sample was $f_{1500}/f_{900}=20.6$. The most
sensitive limit was for C32, in which we could have detected
$f_{1500}/f_{900}=65.9$, while the least sensitive was for
D7, in which we could only have detected a ratio of
$f_{1500}/f_{900}=8.3$. \footnote{The ability to detect larger ratios
of $f_{1500}/f_{900}$ for D7 is limited not only by the object's
fainter-than-average flux at 1500~\AA, but also by 
its relatively low redshift ($z=2.76$), which causes
rest-frame $880-910$~\AA\ to fall at a bluer-than-average observed
wavelength range with higher pixel-to-pixel noise than the 
typical value in our sample.} We therefore conclude that the finite depth of
our observations did not prevent us from detecting objects similar to
D3 and C49. If additional objects with similar $f_{1500}/f_{900}$
had been in the sample, we would have detected Lyman-continuum
flux from them.

Second, in order to determine the range of emergent 
UV-to-Lyman-continuum ratios in the sample,
we must take into account the uncertainty in IGM opacity along
the line of sight in the Lyman-continuum region. 
For the sake of simplicity, though, we first consider the
case in which all objects in the sample are characterized
by the same IGM correction factor, and therefore,
that the relative values (or limits) in $f_{1500}/f_{900}$
translate directly into the relative values
on $f_{1500}/f_{900}$$,_{corr}$. In this case, we
find significant differences among the emergent 
flux-density ratios. Six objects have emergent
UV-to-Lyman-continuum ratios greater than twice
the ratio in C49, while 11 objects have emergent flux-density ratios
greater than twice that in D3. 

Of course, variance in $f_{1500}/f_{900}$ does not necessarily
imply the same level of variance in $f_{1500}/f_{900}$$,_{corr}$.
In qualitative terms, if C49 and D3 happen to have significantly
more transparent lines of sight than the average at $z\sim 3$, while
other objects in the sample with $900$~\AA\ non-detections
have sightlines with significantly
more than average attenuation in the Lyman continuum,
then the non-detections could easily have equivalent
$f_{1500}/f_{900}$$,_{corr}$ values.
At $z\sim 3$, the dispersion in IGM
opacity for individual sightlines is fairly large. 
Once the uncertainty
in IGM opacity for individual sightlines is included,
the constraints are considerably weakened
on the relative emergent flux-density ratios of
C49, D3, and individual Lyman-continuum non-detections in the sample.
In spite of this fact, we can still place some constraints
on the sample variance in $f_{1500}/f_{900}$$,_{corr}$, by
comparing C49 and D3 with the {\it average} of the remaining 12
objects in the sample, for which the observed lower limit on 
the UV-to-Lyman-continuum ratio is $f_{1500}/f_{900} > 43.0$.
This lower limit was estimated from the average
spectrum of the objects without significant Lyman-continuum
detections, using the flux density
at 1500~\AA\ and the $3\sigma$ upper limit at 900~\AA, and 
includes both statistical and systematic uncertainties.
\footnote{The $f_{1500}/f_{900}$ lower
limit may be smaller if the systematic 
flux offset applied to the spectra (see
section~\ref{sec:resultsdetect}) underestimates the true required value.
However, we proceed with the assumption that the offset applied
was the correct one.}
The uncertainty on the average IGM opacity for 12 random sightlines
is significantly smaller than for a single line of
sight. For this 12-spectrum average, 
the 95.4\% confidence interval on the
correction factor for IGM attenuation yields
$f_{1500}/f_{900}$$,_{corr} > 13 - 20$ (the factor 13 [20] corresponds to
the most [least] IGM attenuation), though these numbers do not include
the effects of the SSA22a galaxy overdensity on the Lyman-continuum
opacity.  As discussed above, the 95.4\% confidence limits for
C49 are $f_{1500}/f_{900}$$,_{corr} = 0.3 -  7.5$, 
while they are $f_{1500}/f_{900}$$,_{corr} = 0.2 -  4.6$ for
D3. The lower limits on $f_{1500}/f_{900}$$,_{corr}$
are unphysical, but here we are only interested in
the upper limits. Unless neglecting
the SSA22a galaxy overdensity causes us to underestimate significantly the
average IGM opacity for the 12 non-detections, 
we have indeed demonstrated that
the emergent UV-to-Lyman-continuum ratios in
C49 and D3 are significantly different from the average
of the remaining objects in the sample with only
Lyman-continuum non-detections. The challenge
now is to explain this variation. 

\section{Implications}
\label{sec:implications}

\begin{figure}
\plotone{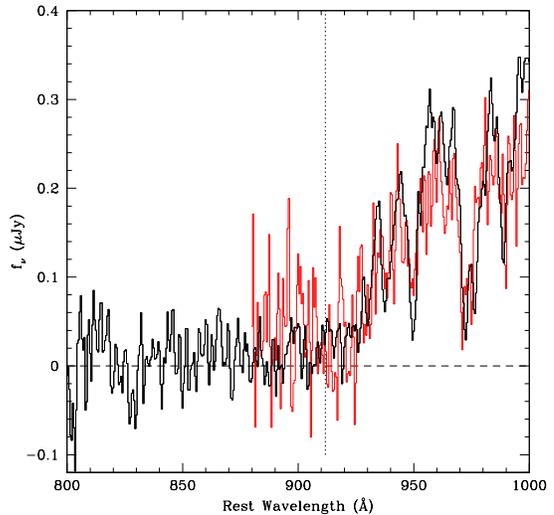}
\caption{A zoomed-in comparison with the results of \citet{steidel2001}.
The composite spectrum from the current sample
is plotted in black, while the \citet{steidel2001}
composite is shown in red. The Lyman limit
is indicated with a vertical dotted line.
The \citet{steidel2001} spectrum has been
scaled to the level of the current composite at 1500~\AA.
The pixel-to-pixel noise level in the Lyman-continuum region of
the \citet{steidel2001} curve is more than
two times larger than in the new composite
spectrum. Over the region $880-910$~\AA,
the average flux density in the \citet{steidel2001}
spectrum is greater than that in the new composite at
high formal significance, despite the level of pixel-to-pixel
noise.
}
\label{fig:lyccontcompzoom}
\end{figure}

In the previous section, we found a sample average 
UV-to-Lyman-continuum flux-density ratio, corrected for IGM absorption, of 
$f_{1500}/f_{900}$$,_{corr} = 22$, which translates to
a relative Lyman-continuum escape fraction of $f_{esc,rel}=14$\%.
The corresponding values in \citet{steidel2001} are
$f_{1500}/f_{900}$$,_{corr} = 4.8$ (slightly different
from the value of $4.6$ presented in that paper, because of the
different method used for correcting for IGM absorption),
and $f_{esc,rel}=63$\%. Therefore, using the average
IGM opacities for the average emission redshifts for
each sample, we find that the estimate
of $f_{esc,rel}$ for the current LBG sample is $\sim 4.5$ times
lower than that of the sample in \citet{steidel2001}. If we consider
the 95.4\% confidence intervals on IGM opacity for 
a sample of 29 sightlines at $\langle z\rangle = 3.40$
and 14 sightlines at $\langle z \rangle = 3.06$, we 
still find that $f_{esc,rel}$ for the present sample
is at least $\sim 3$ times lower than that for
the \citet{steidel2001} sample.
Figure~\ref{fig:lyccontcompzoom} shows a zoomed-in
view of the Lyman-continuum regions for the \citet{steidel2001}
composite (in red) and the new composite (in black),
with the spectra scaled to the same value at
$1500$~\AA. The \citet{steidel2001} spectrum, while noisier, is
still characterized by a mean flux-density value in the 
$880-910$~\AA\ region that
is more than three standard deviations (as measured
by pixel-to-pixel fluctuations) higher
than the mean flux density in our new LBG composite spectrum. 

If we adopt the new estimates for 
$\langle f_{1500}/f_{900}$$,_{corr} \rangle$
and $\langle f_{esc,rel} \rangle$ as representative
of the LBG population, it is possible to scale the values
from \citet{steidel2001}, to estimate the global contributions
of LBGs to the ionizing background at $z \sim 3$.
For this estimate, we use $\langle f_{1500}/f_{900}$$,_{corr} \rangle$
to convert the published LBG $1500$~\AA\ luminosity
function \citep{adelberger2000} to one at $900$~\AA.
This $900$~\AA\ luminosity function is integrated down
to $0.1L^*$ to estimate the comoving emissivity from
LBGs at the Lyman limit.
Taking into account the mean free path of Lyman-continuum
photons at $z\sim 3$, which is $\Delta z \simeq 0.18$,
we find $J_{\nu}(900) \sim  2.6 \times  10^{-22} 
\mbox{ erg s}^{-1}\mbox{cm}^{-2}\mbox{Hz}^{-1}\mbox{sr}^{-1}$. 
\footnote{The \citet{steidel2001} result was derived
using an Einstein-de Sitter cosmology. However, $J_{\nu}$,
which is directly proportional to the product of comoving emissivity and 
Lyman-continuum mean free path, is roughly independent of cosmology. Therefore,
we are justified in applying a simple scaling to the \citeauthor{steidel2001} 
value for $J_{\nu}(900)$.}
If the composite spectrum in \citet{steidel2001}
is adopted as characteristic of the entire
LBG population, the inferred contribution
to the ionizing background is correspondingly higher:
$J_{\nu}(900) \sim  1.2 \times  10^{-21}
\mbox{ erg s}^{-1}\mbox{cm}^{-2}\mbox{Hz}^{-1}\mbox{sr}^{-1}$.
Because the galaxies included in the \citeauthor{steidel2001} composite 
spectrum are drawn only from the bluest quartile of LBGs,
that spectrum may only be representative of $\sim 25$\%
of LBGs. Under the assumption that the bluest quartile
of LBGs are also the most likely to have escaping Lyman-continuum
emission and that the remaining, 
redder, LBG population is characterized by a negligible escape fraction,
the \citeauthor{steidel2001}  composite spectrum translates 
into a $J_{\nu}(900)$ value consistent with the one presented here.
Combining a new measurement of the faint end of the $z\sim3$ QSO
UV luminosity function with existing data on the bright
end, and assuming 100\% escape fraction of ionizing
radiation at all QSO luminosities,
\citet{hunt2004} present estimates of the QSO contribution
to the ionizing background at $z\sim 3$. This estimate
is $J_{\nu}(900) \sim  2.4 \times  10^{-22}
\mbox{ erg s}^{-1}\mbox{cm}^{-2}\mbox{Hz}^{-1}\mbox{sr}^{-1}$,
very similar to the estimate for LBGs presented in this work.
\footnote{The \citet{hunt2004} work also assumed
an Einstein-de Sitter cosmology, and therefore makes for a
fair comparison with the \citet{steidel2001} results, and,
by extension, the current estimate of $J_{\nu}$ from galaxies. 
Because the estimate of $J_{\nu}$ is roughly independent
of cosmology, the values presented for QSOs
and galaxies in \citet{hunt2004} and the current
work, respectively, can also be compared consistently with
$J_{\nu}$ values based on an  
$\Omega_m=0.3$, $\Omega_{\Lambda}=0.7$ cosmology 
\citep[e.g., those in][]{bolton2005}.}
If galaxies and QSOs contribute roughly equal amounts
to the ionizing background at $z\sim 3$, this should be 
reflected in the spectrum of the metagalactic
ionizing radiation field, which in turn determines
the relative rates of ionization of intergalactic hydrogen and helium 
\citep{haardt1996,cen2006,bolton2006}.

It is valuable to compare our results with recent
estimates of the metagalactic hydrogen ionization
rate derived using independent methods. 
Modeling the observed Ly$\alpha$ forest opacity
in QSO absorption spectra using a suite of 
high-resolution hydrodynamical simulations,
\citet{bolton2005} conclude that the global
hydrogen ionization rate at $z\sim 3$ is
$\Gamma_{\mbox{HI}} = 8.6 \pm 3.0 \times 10^{-13} \mbox{s}^{-1}$.
With a spectral slope for $J(\nu)$ of $\nu^{-1.8}$,
as in \citet{madau1999}, this $\Gamma_{\mbox{HI}}$ corresponds to 
$J_{\nu}(900) \sim  3.4 \times  10^{-22}
\mbox{ erg s}^{-1}\mbox{cm}^{-2}\mbox{Hz}^{-1}\mbox{sr}^{-1}$.
If the shape of the ionizing spectrum lies
between $\nu^{-3}$ and $\nu^{0}$, the conversion
between $\Gamma_{\mbox{HI}}$ and $J_{\nu}(900)$
ranges between $4.2$ and $2.2 \times  10^{-22}
\mbox{ erg s}^{-1}\mbox{cm}^{-2}\mbox{Hz}^{-1}\mbox{sr}^{-1}$.
Adopting the \citet{steidel2001} composite spectrum
as representative of the entire LBG population
therefore translates into a LBG contribution
to the ionizing background exceeding
by more than a factor of $\sim 3$
the value of $J_{\nu}(900)$
inferred from modeling the Ly$\alpha$ forest absorption.
In contrast, our new estimate for the contribution 
from LBGs to the ionizing background does not exceed the 
Ly$\alpha$ forest-inferred result, and, furthermore,
the new sum of the contributions from galaxies and QSOs \citep{hunt2004}
is roughly consistent with the total H~I ionizing radiation
field presented in \citet{bolton2005}.
It is worth noting that the quasar proximity effect study
of \citet{scott2000} indicates a value for the ionizing
background at $z\sim 3$ that is roughly twice as 
large as the one in \citeauthor{bolton2005}, and is
arguably a more direct method. The uncertainty
in the quasar proximity effect result is large enough
that the two methods still yield values for $J_{\nu}(900)$ that
are actually statistically consistent
with each other, but it will be important to reconcile
the systematic offset in the mean inferred
$J_{\nu}(900)$ from these different techniques.

Estimating the global contribution from LBGs to the 
ionizing background is based on the assumption that we have determined
the average shape of LBG spectra between the UV and Lyman-continuum
regions. This assumption should be viewed with a degree of
skepticism for more than one reason. First, for the two samples
of LBGs with flux measurements in the Lyman-continuum region,
two different values of $f_{1500}/f_{900}$$,_{corr}$ have been
derived. This fact is not particularly surprising, given that
the two samples are characterized by different average
rest-frame UV spectroscopic properties, which may in turn be related
to the emergent far-UV spectral shape. The \citet{steidel2001}
sample is characterized on average by stronger Ly$\alpha$ emission,
weaker low-ionization interstellar absorption lines, 
and less dust extinction. The weaker interstellar 
absorption and dust extinction may both be correlated
with higher escape fraction of Lyman-continuum
photons in the \citet{steidel2001} sample. Furthermore,
neither of the two Lyman-continuum samples is completely representative
of the total LBG spectroscopic sample. Both the
current and \citeauthor{steidel2001} samples have brighter
than average UV luminosities -- most striking in the current sample, in which
all the galaxies are brighter than $L^*$ of the LBG UV
luminosity function. While the current sample has an average
inferred dust extinction very close to that of the
total LBG sample, the average Ly$\alpha$ emission and interstellar
absorption line strengths are, respectively, weaker and stronger
than in the total LBG sample. While the \citeauthor{steidel2001}
sample has similar average Ly$\alpha$ emission and interstellar
absorption properties to the total LBG sample, it is significantly
bluer in the UV continuum.  Furthermore, most of the galaxies
in the current sample reside in a significant overdensity,
which represents an atypical large-scale environment and
may affect the level of IGM opacity in the Lyman-continuum
region. Therefore, we may have only succeeded
so far in measuring $f_{1500}/f_{900}$$,_{corr}$ for different
types of LBGs drawn from the bright end of the luminosity
function, some of which reside in
a special environment, without much constraint on the average
ratio for the total LBG sample. 

The second reason for skepticism is that, even within our
sample, we have demonstrated a significant variation in 
$f_{1500}/f_{900}$$,_{corr}$. C49 and D3,
the two galaxies for which we detected flux in the 
Lyman-continuum region, must have significantly different
$f_{1500}/f_{900}$$,_{corr}$ ratios from at least
the average of the remaining galaxies in our sample.
The rest-frame UV properties of C49 and D3 do not immediately
stand out, however, relative to those of
the objects without Lyman-continuum detections. 
Furthermore, these two galaxies have
properties strikingly different from each other! Since we do not yet
have a coherent physical explanation --  in terms of stellar population
age or ISM geometry --  for why these two galaxies alone
exhibit significant flux in the Lyman-continuum region, 
we are unable to show that our small sample is representative even of 
bright LBGs. 
As the intrinsic $(L_{1500}/L_{900})_{int}$ ratio depends
so sensitively on stellar population age, it is important
to obtain near- and mid-IR data, which will allow us to constrain
the stellar populations of these galaxies. 
On the other hand, the variation in
$f_{1500}/f_{900}$$,_{corr}$ among our sample may simply reflect the
fact that the escape of Lyman-continuum emission is
an anisotropic process, resulting in a complex spatial
distribution of Lyman-continuum emission relative to that
of the UV-continuum emission determining the positions
of our spectroscopic slits.

What is required is a statistical sample of LBGs with
deep Lyman-continuum measurements. This sample should be roughly an
order of magnitude larger than the current one, 
span a larger range in luminosity, with multi-wavelength data 
from which constraints on stellar population ages can be obtained.
Furthermore, to enable robust correction for IGM
opacity along the line of sight, the future sample must contain objects probing
at least several statistically independent cosmic volumes -- as opposed 
to the objects in this work, all of which are contained 
within $9\times13$ comoving Mpc$^2$ on the sky and most 
within a previously-discovered galaxy overdensity
\citep{steidel1998}. This larger sample should also contain a more significant
set of Lyman-continuum detections, whose properties 
can be characterized on a statistical basis, relative
to those of objects without detections.  A lack of intrinsic difference
between the two populations would support the idea that variation
in $f_{1500}/f_{900}$$,_{corr}$ reflects anisotropic Lyman-continuum
emission being sampled with different degrees of efficiency.
Only with a survey satisfying
the above requirements will we be truly justified in translating the LBG
UV luminosity function into a constraint on the global
contribution of LBGs to the ionizing background. However,
what we have shown here, the first detection of ionizing radiation
from {\it individual} LBGs at $z\sim 3$, represents
significant progress in the right direction.

\bigskip
We would like to thank our collaborators in the LBG survey
for their assistance in various stages of the project,
and an anonymous referee, whose comments improved the paper.
We wish to extend special thanks to those of Hawaiian ancestry on
whose sacred mountain we are privileged to be guests. Without their generous
hospitality, most of the observations presented herein would not
have been possible. We also thank the staff
at the W.~M. Keck Observatory for their 
assistance with the LRIS observations.
Finally, we gratefully acknowledge Michael 
Santos and Jerry Ostriker for extensive and enlightening discussions,
and Robert Lupton for statistical insights.
CCS and DKE have been supported by grants
AST00-70773 and AST03-07263 from the U.S. National Science 
Foundation and by the David and Lucile
Packard Foundation. AES acknowledges
support from the Miller Foundation for Basic Research in Science.


\clearpage
\clearpage
\begin{landscape}
\begin{deluxetable}{cllccrcccc}
\small
\tablewidth{0pc}
\footnotesize
\tabletypesize{\footnotesize}
\tablecaption{LRIS Observations of Bright Lyman-Break Galaxies in the SSA22a Field\label{tab:obs}}
\tablehead{
\colhead{} & \colhead{} & \colhead{} & \colhead{} & \colhead{} & \colhead{} &
 \colhead{} & \colhead{} &
\colhead{LRIS-B} & \colhead{LRIS-R}  \\
\colhead{Object} & \colhead{R.A. (J2000)} &
\colhead{Dec. (J2000)} &
\colhead{${\cal R}$\tablenotemark{a}} &
\colhead{$G-{\cal R}$\tablenotemark{a}} &
\colhead{$U_n-G$\tablenotemark{a}} &
 \colhead{$z_{em}$\tablenotemark{b}} & \colhead{$z_{abs}\tablenotemark{c}$} &
\colhead{Exposure (s)\tablenotemark{d}} & \colhead{Exposure (s)} }
\startdata
D17 & 22:17:18.88 & 00:18:16.89 & 24.27 & 0.45 & $2.01$ & 3.0851,3.0945 & 3.0697 & 29200/69040 & 61000 \\ 
C24 & 22:17:18.94 & 00:14:45.65 & 23.86 & 0.78 & $>2.77$ & 3.1026  & 3.0960 & 29200/79240 & 61000 \\ 
C49 & 22:17:19.81 & 00:18:18.64 & 23.85 & 0.59 & $>3.04$ & 3.1601,3.1656 & 3.1492 & 29200/58040 & 61000 \\ 
C35 & 22:17:20.23 & 00:16:52.33 & 24.18 & 0.95 & $>2.52$ & \nodata  & 3.0980 & 29200/79240 & 61000 \\ 
C47 & 22:17:20.23 & 00:17:32.45 & 23.84 & 0.60 & $>3.14$ & 3.0748  & 3.0651 & 29200/79240 & 61000 \\ 
C41 & 22:17:24.45 & 00:17:14.75 & 23.80 & 0.18 & $>3.49$ & 3.0236  & 3.0166 & 29200/79240 & 61000 \\ 
C32 & 22:17:25.62 & 00:16:13.10 & 23.68 & 0.67 & $>3.19$ & 3.2895,3.2991 & 3.2921 & 29200/79240 & 61000 \\ 
C11 & 22:17:25.68 & 00:12:35.30 & 24.20 & 0.47 & $>2.95$ & 3.1014  & 3.0962 & 29200/79240 & 61000 \\ 
MD23 & 22:17:28.01 & 00:14:29.73 & 24.14 & 0.46 & $1.92$ & 3.0845  & 3.0753 & 29200/79240 & 61000 \\ 
D7 & 22:17:28.79 & 00:12:47.07 & 23.50 & 0.62 & $2.14$ & \nodata  & 2.7564 & 29200/47840 & 61000 \\ 
C16 & 22:17:31.96 & 00:13:16.11 & 23.64 & 0.98 & $>2.88$ & \nodata  & 3.0651 & 29200/79240 & 61000 \\ 
D3-ap1 & 22:17:32.42 & 00:11:32.97 & 23.37 & 0.97 & $2.58$ & 3.0716  & 3.0649 & 29200/47840 & 29200 \\ 
D3-ap2 & 22:17:32.42 & 00:11:32.97 & 23.37 & 0.97 & $2.58$ & 3.0767  & 3.0677 & 29200/47840 & 29200 \\ 
MD14 & 22:17:37.93 & 00:13:44.21 & 24.14 & 0.86 & $2.25$ & \nodata  & 3.0969 & 29200/29200 & 29200 \\ 
C22 & 22:17:38.18 & 00:14:32.06 & 24.46 & 0.54 & $>2.72$ & \nodata  & 2.8765 & 29200/58040 & 61000 \\ 
\enddata
\tablenotetext{a}{AB magnitude system;
effective wavelengths are 4730, and 6830 \AA\, for
$G$ and ${\cal R}$, respectively
\citep[see also][]{steidel2003}.}
\tablenotetext{b}{Spectroscopic redshift measured from the observed wavelength
of Ly$\alpha$ emission. Three objects, D17, C49, and C32, exhibit
double-peaked Ly$\alpha$ profiles. For these objects, we list the emission
redshifts corresponding to both Ly$\alpha$ peaks.}
\tablenotetext{c}{Average spectroscopic redshift measured from the observed 
wavelengths of strong low-ionization interstellar absorption lines.}
\tablenotetext{d}{$\:$ LRIS-B exposure times are listed for
$\lambda \leq 4000$~\AA\ (left) and $4000$~\AA $\leq \lambda \leq 5000$~\AA\ 
(right). }
\end{deluxetable}
\clearpage
\end{landscape}

\begin{deluxetable}{lccc}
\tablewidth{0pc}
\footnotesize
\tablecaption{Properties of Different LBG Samples}
\tablehead{
\colhead{} & \colhead{This Work}  & \colhead{Total LBG Sample} & \colhead{\citet{steidel2001} Sample} 
}
\startdata
$N_{gal}$ & 14 &  811 & 29\\
$\langle z \rangle$ & $3.06\pm0.09$ &  $2.96\pm0.29$ & $3.40\pm0.18$   \\
$\langle {\cal R}\rangle$& $23.92\pm0.32$ & $24.60\pm0.56$ & $24.34\pm0.38$ \\
$\langle M_{UV}\rangle$\tablenotemark{a} & $-21.63\pm0.32$ & $-20.88\pm0.58$ & $-21.40\pm0.38$ \\
$\langle G-{\cal R} \rangle$\tablenotemark{b} & $0.65 \pm 0.23$ & $0.61\pm 0.29$ & $0.89\pm 0.18$ \\
$\langle E(B-V) \rangle$\tablenotemark{c} & $0.11\pm 0.08$  & $0.13\pm  0.09$ & $0.07\pm 0.05$ \\
$\langle W_0(\mbox{Ly}\alpha)\rangle$\tablenotemark{d} & 1.7 \AA\  &  15.2 \AA\ & 14.1 \AA\ \\
$\langle W_0(IS) \rangle$\tablenotemark{d} & $-2.2$ \AA\  & $-1.8$ \AA\ & $-1.7$ \AA\ \\
\enddata
\tablenotetext{a}{AB absolute $UV$ magnitude, derived from the
apparent ${\cal R}$ magnitude. 
At $z\sim 3.0-3.4$, the
${\cal R}$-band effective wavelength corresponds to $1550-1700$~\AA.}
\tablenotetext{b}{The larger $<G-R>$ 
of the \citet{steidel2001} sample reflects the increased
average IGM opacity at $z=3.40$ relative to
$z=3.06$ and $z=2.96$. In fact, this sample appears
to have bluer {\it intrinsic} colors on average
than the two other samples.
}
\tablenotetext{c}{Dust extinction, parametrized by $E(B-V)$,
is derived from the observed $G-R$ color corrected for the effects of intergalactic absorption and
intrinsic Ly$\alpha$ emission/absorption. The corrected $G-R$ color
is modeled, and $E(B-V)$ inferred, using an unreddened \citet{bc2003} 
300~Myr continuous star-formation model template 
spectrum and a \citet{calzetti2000} attenuation curve  
\citep[see][]{steidel1999,adelberger2000,shapley2001}.
}
\tablenotetext{d}{ Rest-frame Ly$\alpha$ and low-ionization interstellar
absorption equivalent widths have been measured from the average composite
spectrum for each sample. Positive numbers refer to 
emission and negative numbers to absorption. In the case of Ly$\alpha$, the
reported equivalent width represents the total profile, including
both emission and absorption components. 
For both the Total LBG and \citet{steidel2001}
samples, the total and emission profiles have very similar equivalent
widths. In the case of the current sample, the total equivalent width
is close to $\sim 0 $~\AA\, while the emission equivalent width
is $\sim 5$~\AA\ rest. The reported low-ionization interstellar absorption
equivalent width for each sample represents the average of the four strongest features:
Si~II $\lambda 1260$, O~I $\lambda 1302$ + Si~II $\lambda 1304$,
C~II $\lambda 1334$, and Si~II $\lambda 1526$.
}
\end{deluxetable}

\clearpage

\end{document}